\documentclass[11pt,a4]{article}
\usepackage{amssymb}
\usepackage{amsmath}

\input{tcilatex}
\begin{document}

\title{Abelian Magnetic Monopoles and Topologically Massive Vector Bosons in
Scalar-Tensor Gravity with Torsion Potential}
\author{S. A. Ali$^1$, C. Cafaro$^1$, S. Capozziello$^2$, Ch. Corda$^3$ \\
{\small \textit{$^1$Department of Physics, State University of New York at
Albany,}}\\
{\small \textit{1400 Washington Avenue, Albany, NY 12222, USA}}\\
{\small \textit{$^2$Dipartimento di Scienze Fisiche, Universit\`a di Napoli
{}`` Federico II''and INFN Sez. di Napoli,}}\\
{\small \textit{Compl. Univ. di Monte S. Angelo, Edificio G, Via Cinthia,
I-80126, Napoli, Italy}}\\
{\small \textit{$^3$Universit\`a di Pisa and INFN Sez. di Pisa,}} \\
{\small \textit{Via F. Buonarroti 2,I-56127, Pisa, Via E. Amaldi, I-56021
Pisa, Italy}}}
\maketitle

\begin{abstract}
A Lagrangian formulation describing the electromagnetic interaction - \
mediated by topologically massive vector bosons - between charged, spin-$%
\frac{1}{2}$ fermions with an abelian magnetic monopole in a curved
spacetime with non-minimal coupling and torsion potential is presented. The
covariant field equations are obtained. The issue of coexistence of massive
photons and magnetic monopoles is addressed in the present framework. It is
found that despite the topological nature of photon mass generation in
curved spacetime with isotropic dilaton field, the classical field theory
describing the nonrelativistic electromagnetic interaction between a
point-like electric charge and magnetic monopole is inconsistent.

\end{abstract}

\section{Introduction}

Extended Theories of Gravity have become a sort of paradigm in the study of
gravitational interaction since several motivations push for enlarging the
traditional scheme of Einstein's General Relativity (GR) \cite{GRG}. Such
issues come, essentially, from cosmology and quantum field theory. In the
first case, it is well known that higher-order derivative theories and
scalar-tensor theories give rise to inflationary cosmological solutions
capable, in principle, of solving the shortcomings of the Standard
Cosmological Model. Besides, they have relevant features also from the
quantum cosmology viewpoint. In the second case, every unification scheme as
Superstrings, Supergravity or Grand Unified Theories, takes into account
effective actions where nonminimal couplings to the geometry or higher-order
terms in the curvature invariants come out. Such contributions are due to
one-loop or higher-loop corrections in the high-curvature regimes near the
full (not yet available) quantum gravity regime. In the weak-limit
approximation, all these classes of theories should be expected to reproduce
Einstein's GR which, in any case, is experimentally tested only in this
limit. This issue is debatable however, since several relativistic theories
do not reproduce those of GR in the Newtonian approximation.

Magnetic monopoles were first proposed by Dirac in the framework of
classical electrodynamics in his classic works \cite{Dirac}. The main
purpose for the introduction of monopoles was to provide a physical
explanation for the quantization of electric charge. This is known as the
Dirac quantization rule. Antisymmetric tensor gauge fields analogous to the
torsion potential employed here were proposed some time ago in the
literature. The first description of such an antisymmetric field was due to
Ogievetskii and Polubarinov \cite{Ogievetskii}. In 1973, Kalb and Raymond 
\cite{KalbRaymond} described classical string interactions by means of an
antisymmetric field the interpretation of which is that of a potential
generated by the string. Scherk and Schwarz \cite{ScherkSchwarz} showed that
torsion could be viewed as the product of the antisymmetric field of string
theory multiplied by a scalar field. In their work, the spacetime metric was
not covariantly constant. In \cite{FradkinTseytlin}, Fradkin and Tseytlin
derived an effective Lagrangian density in the low energy limit of string
theory, describing not only gravity but also a scalar (dilaton) and
antisymmetric field.

In the present work we implement a scalar-tensor generalization of gravity
in the sense of Brans-Dicke \cite{BransDicke, Brans} with non-vanishing
curvature and torsion, whereby the gravitational coupling constant becomes a
scalar field. This scalar field is identified as the dilaton. Being a new
dynamical variable of the theory, we include a kinetic term for the dilaton
in the total system Lagrangian density. Gravitational theories with torsion
such as Einstein-Cartan theory \cite{Hehl1} or Poincar\'{e} gauge theory 
\cite{Kibble, cho2, Blagojevic, Grignani} describe the torsion as being the
anti-symmetric part of a generalized affine connection or as the Cartan
structure equation for the dual frame field taken as a gauge potential. By
contrast, we assume in this paper that torsion is derived from a
(anti-symmetric) second-rank tensor potential \cite{Hammond1} which could be
further generalized by considering bi-vectors \cite{noi}. We admit
topological interaction between torsion and electromagnetic gauge potentials
and consider the electromagnetic interaction between charged,
nonrelativistic spin-$\frac{1}{2}$ fermions with an abelian magnetic
monopole. As a consequence of the electromagnetic and torsion gauge field
coupling, the fermion-monopole interaction is mediated by topologically
massive vector bosons. The material Lagrangian density is taken to be that
of the Dirac minimally coupled type.

It is known that massive photons and magnetic monopoles of the Dirac type
cannot coexist within the same theory defined over flat Minkowski spacetime 
\cite{joshi, cafali}. In such scenarios, the photon mass is usually
introduced in an \textit{ad hoc} manner by explicitly breaking the gauge
symmetry of the theory. In this work we consider whether such an
incompatibility emerges from the a priori inclusion of photon mass or from
the specific mechanism for gauge boson mass generation. Moreover, due
attention is given to the role of the curved spacetime geometry and
isotropic dilation field with regard to this incompatibility.

The paper is organized as follows: In Section 2, the Lagrangian density
representing the matter background is specified. In Section 3, the total
system Lagrangian density including gravity, gauge, matter and interaction
terms is obtained. The electromagnetic and torsion gauge field, Einstein,
Klein-Gordon, (nonlinear) Dirac equations of Heisenberg-Pauli type and
Bianchi identities in the electromagnetic and torsion sectors are obtained
in Section 4. In Section 5, we investigate the possibility of coexistence of
magnetic monopoles with topologically massive vector bosons within the
framework of Scalar-Tensor Gravity with Torsion Potential. Our conclusions
are presented in Section 6.

\section{The Matter Background}

To make a distinction between the coordinate (or holonomic indices) and
local Lorentz (or non-holonomic) coordinates, we use Greek indices $(\mu $, $%
\nu =0$, $1$, $2$, $3)$ for the former and Latin indices $(j$, $k=0$, $1$, $2
$, $3)$ for the latter. Latin indices are raised and lowered with the
Minkowski metric $\eta _{ij}$. Greek indices are raised and lowered with the
spacetime metric $g_{\alpha \beta }$. We use geometrized units ($\hbar =c=1$%
) throughout this work.

In the present Section, we are concerned with constructing a Lagrangian
formulation of the dynamics of spinor valued fields\textbf{\ }$\psi (x)$%
\textbf{\ }defined over a curved manifold endowed with torsion. The
equations of motion are given as the Euler-Lagrange equations for the
corresponding action-integral $I\left( \Omega \right) =\int_{\Omega }d^{4}x%
\mathcal{L}\left( \psi \left( x\right) \text{, }\partial \psi \left(
x\right) \text{; }x\right) $ defined over a spacetime volume $\Omega $. In
order to introduce spinor fields in the Riemann-Cartan geometry considered
here, it is convenient to choose an orthonormal (Lorentz) basis vectors $%
e_{i}=e_{i}^{\alpha }\left( x\right) e_{\alpha }$\textbf{\ }for the tangent
space satisfying\textbf{\ }$e_{i}\cdot e_{j}=\eta _{ij}$ where\textbf{\ }$%
e_{\alpha }\equiv \partial _{\alpha }=\frac{\partial }{\partial x^{\alpha }}$%
\textbf{\ }represents a coordinate basis in the tangent space $T_{P}$\textbf{%
\ }at point $P$\textbf{\ }in the spacetime manifold,\textbf{\ }$\eta _{ij}=$%
diag$(-1$, $+1$, $+1$, $+1)$ is the Minkowski metric and $g_{\alpha \beta
}=e_{\alpha }\cdot e_{\beta }$ is the metric of curved spacetime\textbf{. }%
The metrics $\eta _{ij}$ and $g_{\alpha \beta }$ are related via%
\begin{equation}
g_{\alpha \beta }:=e_{\alpha }^{\;i}\left( x\right) e_{\beta }^{\;j}\left(
x\right) \eta _{ij}\text{.}  \label{metric}
\end{equation}%
The quantities $e_{\sigma }^{\;i}\left( x\right) $, called tetrads, are
coefficients of the dual ($1$-form) basis non-holonomic co-vectors $%
\vartheta ^{a}\left( x\right) =e_{\text{ }\gamma }^{a}(x)dx^{\gamma }$ that
satisfy the orthogonality relations $e_{\text{ }j}^{\alpha }e_{\text{ }%
\alpha }^{i}=\delta _{\text{ }j}^{i}$. In particular, the tetrads constitute
transformation matrices that map from local Lorentz (with non-holonomic
coordinates $x^{a}$) to coordinate (with holonomic coordinates $x^{\mu }$)
bases, i.e., $v^{\alpha }=e_{i}^{\alpha }v^{i}$ with $v^{i}=v^{\alpha
}e_{\alpha }^{i}$. The components $e_{\alpha }^{\;i}\left( x\right) $ and $%
e_{i}^{\;\alpha }\left( x\right) $ transform as covariant and contravariant
vectors (under the Poincar\'{e} group) of the frame $x^{\mu }$, if and only
if the rotations $\partial _{\lbrack \mu }e_{\lambda ]}^{\;a}$\ vanish at
all points. The equations $\partial _{\lbrack \mu }e_{\lambda ]}^{\;i}=0$\
are the so-called integrability conditions \cite{Kibble}, implying $%
e_{\lambda }^{\;i}\left( x\right) =\partial _{\lambda }x^{i}$. If the
integrability condition is not satisfied, the reference frame formed by $%
e_{i}^{\;\beta }\left( x\right) $\ and $e_{\lambda }^{\;i}\left( x\right) $
is said to be non-holonomic. The quantities 
\begin{equation}
\Omega _{cab}:=e_{\nu c}(x)\left[ e_{\text{ \ }a}^{\mu }(x)\partial _{\mu
}e_{\text{ \ }b}^{\nu }(x)-e_{\text{ \ }b}^{\mu }(x)\partial _{\mu }e_{\text{
\ }a}^{\nu }(x)\right] \text{,}
\end{equation}%
are called the objects of non-holonomicity. They measure the
non-commutativity of the tetrad basis \cite{Hehl1}. We may readily define
tensors and various algebraic operations with tensors at a given point in
the spacetime manifold. Comparison of tensors at different points however,
requires introduction of a linear connection via the process of parallel
transport. The linear connection defines a covariant derivative operator%
\textbf{\ }$\hat{D}$\textbf{.} In non-holonomic coordinates, the parallel
transport of an orthonormal basis $e_{i}$\textbf{\ }is given by \cite{Hehl1}%
\textbf{\ }$\delta e_{i}=-\omega _{\text{ }ij}^{k}e_{\alpha
}^{j}e_{k}dx^{\alpha }=-\omega _{\text{ }i\alpha }^{k}e_{k}dx^{\alpha }$.%
\textbf{\ }The associated covariant derivative is given by\textbf{\ }$D_{\mu
}\psi =\left( \partial _{\mu }+\omega _{\mu }^{ab}\gamma _{ab}\right) \psi $%
, while for the contravariant components of a non-holonomic vector we have $%
D_{\mu }v^{i}:=\partial _{\mu }v^{i}+\omega _{\text{ }j\mu }^{i}v^{j}$%
\textbf{. }Note that\textbf{\ }$D_{\mu }$\textbf{\ }is a coordinate
representation of the operator\textbf{\ }$\hat{D}$\textbf{. }The coefficients%
\textbf{\ }$\omega _{ab\mu }$ are known as the spin-connection and\textbf{\ }%
the matrix $\gamma _{ik}$\ is an irreducible spinoral representations of the
Lorentz group defined by%
\begin{equation}
\gamma _{ik}=\frac{1}{2}\left( \gamma _{i}\gamma _{k}-\gamma _{k}\gamma
_{i}\right) \text{.}
\end{equation}%
\ Under local Lorentz transformation (LT) the covariant derivative itself
should transform as a scalar since it does not carry a Lorentz (Latin)
index. Thus\textbf{\ }$D_{\mu }v^{i}\overset{\text{LT}}{\rightarrow }D_{\mu
}^{\prime }v^{\prime i}=\Lambda _{j}^{i}D_{\mu }v^{j}$ where $\Lambda
_{j}^{i}:=\frac{\partial x^{i}}{\partial x^{j}}$ is a non-holonomic
transformation matrix. Making use of the equation for\textbf{\ }$D_{\mu
}v^{i}$,\textbf{\ }$D_{\mu }^{\prime }v^{\prime i}$ and the fact that\textbf{%
\ }$\partial _{\mu }\eta _{ab}=0$ (since the Minkowski metric is constant)
we obtain the transformation property of the spin connection 
\begin{equation}
\omega _{\text{ \ \ }\mu }^{ab}\rightarrow \omega _{\text{ \ \ }\mu
}^{\prime ab}=\Lambda _{i}^{a}\Lambda _{j}^{b}\omega _{\text{ \ \ }\mu
}^{ij}-\left( \partial _{\mu }\Lambda _{i}^{a}\right) \Lambda ^{bi}\text{.}
\label{spin-trans}
\end{equation}%
Parallel transport is a unique geometric operation that is independent of
the choice of frame. The relative rotation of a coordinate (holonomic) basis
vector\textbf{\ }$e_{\alpha }$\textbf{\ }is given by\textbf{\ }$dx^{\alpha
}\left( \partial _{\alpha }e_{k}^{\text{ }\gamma }+\Gamma _{\alpha \beta }^{%
\text{ \ \ \ }\gamma }e_{k}^{\text{ }\beta }\right) e_{\gamma }=dx^{\alpha
}\left( \nabla _{\alpha }e_{k}^{\text{ }\beta }\right) e_{\beta }^{\text{ }%
j}e_{j}$ with the affine\textbf{\ }connection $\Gamma _{\;\mu \nu }^{\rho
}=e_{i}^{\text{ \ }\rho }\left( x\right) D_{\nu }e_{\text{ \ }\mu
}^{i}\left( x\right) =-e_{\mu }^{\text{ \ }i}\left( x\right) D_{\nu }e_{%
\text{ \ }i}^{\rho }\left( x\right) $\textbf{\ }defining the covariant
derivative $\nabla _{\alpha }:=\partial _{\alpha }+\Gamma _{\alpha }^{\beta
\gamma }\Xi _{\beta \gamma }$\textbf{. }The matrices $\Xi _{\alpha \beta
}=-\Xi _{\beta \alpha }$\ are generators of the Lorentz group satisfying the
Lie algebra%
\begin{equation}
\lbrack \Xi _{ij}\text{, }\Xi _{kl}]=\eta _{ik}\Xi _{jl}+\eta _{jl}\Xi
_{ik}-\eta _{jk}\Xi _{il}-\eta _{il}\Xi _{jk}\text{,}  \label{LorentzGen}
\end{equation}%
with $\Xi _{ij}=e_{i}^{\alpha }e_{j}^{\beta }\Xi _{\alpha \beta }$. At this
juncture we emphasize that there is only one linear connection. It may be
expressed in either holonomic or non-holonomic frames of reference. As will
be shown, these two representations of the linear connection are related by (%
\ref{gamma_omega}). Moreover, the linear connection (expressed in either
reference frame) is not a priori torsion free. Indeed, it will be shown that
the linear connection does contain torsion, the latter being equivalently
defined by either (\ref{inter5}) or (\ref{torsion}).

The covariant derivative of a quantity\textbf{\ }$v^{\lambda }$\textbf{\ (}$%
v_{\gamma }$\textbf{) }which behaves like a contravariant (covariant) vector
under the local Poincar\'{e} transformation is given by 
\begin{equation}
\nabla _{\nu }v^{\lambda }=\partial _{\nu }v^{\lambda }+\Gamma ^{\lambda
}\,_{\mu \nu }v^{\mu }\text{, \ }\nabla _{\nu }v_{\mu }=\partial _{\nu
}v_{\mu }-\Gamma ^{\lambda }\,_{\mu \nu }v_{\lambda }\text{.}
\end{equation}%
In analogy to (\ref{spin-trans}), the transformation property for the affine
connection coefficients $\Gamma _{\;\mu \nu }^{\rho }$ is given by%
\begin{equation}
\Gamma _{\text{ \ }\mu \nu }^{\lambda }\rightarrow \Gamma _{\text{ \ }\mu
\nu }^{\prime \lambda }=\Lambda _{\text{ \ }\mu }^{\alpha }\Lambda _{\text{
\ }\nu }^{\beta }\Lambda _{\gamma }^{\text{ \ }\lambda }\Gamma _{\text{ \ }%
\alpha \beta }^{\gamma }+\Lambda _{\text{ \ }\mu }^{\alpha }\Lambda _{\rho
}^{\text{ \ }\lambda }\Lambda _{\text{ \ }\alpha \nu }^{\rho }\text{,}
\label{affine-trans}
\end{equation}%
where $\Lambda _{\text{ \ }\mu }^{\alpha }:=\frac{\partial x^{\alpha }}{%
\partial x^{\mu }}$ is the holonomic transformation matrix and $\Lambda _{%
\text{ \ }\alpha \nu }^{\rho }\equiv \partial _{\alpha }\partial _{\nu
}x^{\rho }$. In view of the inhomogenous term $\Lambda _{\text{ \ }\mu
}^{\alpha }\Lambda _{\rho }^{\text{ \ }\lambda }\Lambda _{\text{ \ }\alpha
\nu }^{\rho }$ in (\ref{affine-trans}), the linear connection is not a
tensor.

The parallel transport of a vector around an infinitesimal closed path is
proportional to the curvature of the manifold which may be calculated as 
\cite{Schouten}%
\begin{equation}
\left[ D_{k}\text{, }D_{l}\right] \psi \left( x\right) =\frac{1}{2}%
R^{ij}{}_{kl}\gamma _{ij}\psi \left( x\right) +C_{kl}^{i}D_{i}\psi \left(
x\right) \text{,}  \label{defalg}
\end{equation}%
with $D_{k}:=e_{k}^{\mu }D_{\mu }$. The central charge $R^{ij}{}_{kl}$ and
structure functions $C_{\text{ }jk}^{i}$ of the deformed algebra (\ref%
{defalg}) are given (in non-holonomic coordinates) by the Cartan structure
equations%
\begin{equation}
R_{\text{ \ }kl}^{ij}\left( \omega \right) :=e_{\lambda }^{i}e^{j\rho }R_{%
\text{ }\rho \mu \nu }^{\lambda }=\partial _{\mu }\omega _{\text{ \ }\nu
}^{ij}-\partial _{\nu }\omega _{\text{ \ }\mu }^{ij}+\omega _{\text{ }k\mu
}^{i}\omega _{\text{ \ \ }\nu }^{kj}-\omega _{\text{ }k\nu }^{i}\omega _{%
\text{ \ \ }\mu }^{kj}\text{, }C_{\text{ }jk}^{i}=\left( e_{\text{ \ }%
j}^{\mu }e_{\text{ \ }k}^{\nu }-e_{\text{ \ }k}^{\mu }e_{\text{ \ }j}^{\nu
}\right) D_{\nu }e_{\mu }^{\text{ \ }i}\left( x\right) \text{.}
\label{struct}
\end{equation}%
The curvature tensor\textbf{\ }$R_{\text{ }\rho \mu \nu }^{\lambda }$\textbf{%
\ }(expressed in holonomic coordinates) is defined by,%
\begin{equation}
R_{\text{ \ }\gamma \rho \lambda }^{\alpha }\left( \Gamma \right) =\partial
_{\gamma }\Gamma _{\rho \lambda }^{\alpha }-\partial _{\rho }\Gamma _{\gamma
\lambda }^{\alpha }+\Gamma _{\gamma \sigma }^{\alpha }\Gamma _{\rho \lambda
}^{\sigma }-\Gamma _{\rho \sigma }^{\alpha }\Gamma _{\gamma \lambda
}^{\sigma }\text{.}  \label{curvature}
\end{equation}%
It is interesting to observe the similarity in structure of the curvature
tensors in (\ref{curvature}) and the first equation in (\ref{struct}).
Indeed, there is only one curvature tensor since these two quantities can be
transformed into each other via appropriate tetrad index saturation,\textbf{%
\ }$R_{\text{ \ }jkl}^{i}\left( \omega \right) =e_{\alpha }^{i}e_{j}^{\gamma
}e_{k}^{\rho }e_{l}^{\lambda }R_{\text{ \ }\gamma \rho \lambda }^{\alpha
}\left( \Gamma \right) $\textbf{. }We can therefore view\textbf{\ }$R_{\text{
\ }\gamma \rho \lambda }^{\alpha }\left( \Gamma \right) $ \textbf{in} (\ref%
{curvature}) and\textbf{\ }$R_{\text{ \ }kl}^{ij}\left( \omega \right) $%
\textbf{\ }in\textbf{\ }(\ref{struct}) as holonomic and non-holonomic
representations, respectively, of the same spacetime curvature.

Since the basis vectors (in either holonomic or non-holonomic frames) change
from one point in the spacetime manifold to another, the derivative of a
vector must be given by \cite{crawford} $\partial _{\mu }v=\partial _{\mu
}\left( v^{i}e_{i}\right) =\left( \partial _{\mu }v^{i}\right)
e_{i}+v^{i}\left( \partial _{\mu }e_{i}\right) \equiv \left( \nabla _{\mu
}v^{i}\right) e_{i}$.\ This implies that $\partial _{\mu }e_{j}=\omega _{%
\text{ \ }j\mu }^{i}e_{i}$. For similar reasons, we conclude $\partial _{\mu
}e_{\nu }=\Gamma _{\text{ \ }\nu \mu }^{\rho }e_{\rho }$. Thus, if we choose
a transformation in (\ref{spin-trans}) which leads from a non-holonomic to a
holonomic frame, then we find \cite{Hehl1, crawford}%
\begin{equation}
\partial _{\nu }e_{i}\,^{\lambda }-\omega ^{k}\,_{i\nu }e_{k}\,^{\lambda
}+\Gamma ^{\lambda }\,_{\mu \nu }e_{i}\,^{\mu }\equiv \mathcal{D}_{\nu
}e_{i}\,^{\lambda }=0\text{, }\partial _{\nu }e^{i}\,_{\mu }+\omega
^{i}\,_{k\nu }e^{k}\,_{\mu }-\Gamma ^{\lambda }\,_{\mu \nu }e^{i}\,_{\lambda
}\equiv \mathcal{D}_{\nu }e^{i}\,_{\mu }=0\text{,}  \label{inter6a}
\end{equation}%
since $\partial _{\mu }e_{j\nu }=\partial _{\mu }\left( e_{j}\cdot e_{\nu
}\right) =\omega _{\text{ \ }j\mu }^{i}e_{i}\cdot e_{\nu }+\Gamma _{\text{ \ 
}\nu \mu }^{\rho }e_{j}\cdot e_{\rho }=\omega _{\text{ \ }j\mu }^{i}e_{i\nu
}+\Gamma _{\text{ \ }\nu \mu }^{\rho }e_{j\rho }$. Observe that $\mathcal{D}%
_{\nu }=\mathcal{D}_{\nu }\left( \Gamma +\omega \right) $. Recalling (\ref%
{metric}) and using (\ref{inter6a}), we may derive the so-called metricity
condition $\nabla _{\lambda }\left( \Gamma \right) g_{\mu \nu }=\mathcal{D}%
_{\lambda }\left( \Gamma +\omega \right) g_{\mu \nu }=\mathcal{D}_{\lambda
}\left( \Gamma +\omega \right) \left( e_{\mu }^{\;i}\left( x\right) e_{\nu
}^{\;j}\left( x\right) \eta _{ij}\right) =0$. This metricity condition
enables the definition of the linear connection\textbf{\ }$\Gamma _{\rho \mu
}^{\sigma }=\mathring{\Gamma}_{\rho \mu }^{\sigma }+T_{\rho \mu }^{\sigma }$%
\textbf{, }where the quantity\textbf{\ }$\mathring{\Gamma}_{\text{ \ }\rho
\mu }^{\sigma }$\textbf{\ }can be identified as the Christoffel connection
coefficient%
\begin{equation}
\mathring{\Gamma}_{\rho \mu }^{\sigma }:=\frac{1}{2}g^{\kappa \sigma }\left(
\partial _{\kappa }g_{\rho \mu }+\partial _{\rho }g_{\mu \kappa }-\partial
_{\mu }g_{\kappa \rho }\right)   \label{Christoffel}
\end{equation}%
and\textbf{\ }$T_{\rho \mu }^{\sigma }$\textbf{\ }is the torsion tensor
defined as the asymmetric part of the affine connection,%
\begin{equation}
T_{\text{ }\beta \gamma }^{\alpha }:=\Gamma _{\text{ }\beta \gamma }^{\alpha
}-\Gamma _{\text{ }\gamma \beta }^{\alpha }\text{.}  \label{inter5}
\end{equation}%
With (\ref{Christoffel}) and (\ref{inter5}) in hand, the quantity $R_{\text{
\ }ijl}^{k}$ in (\ref{struct}) can be expressed in terms of its torsion-free 
$\mathring{R}_{\text{ \ }ijl}^{k}$ and torsion dependant contributions as 
\cite{Schouten}\textbf{\ }%
\begin{equation}
R_{\text{ \ }ijl}^{k}=e_{l}^{\lambda }\left( x\right) e_{\alpha }^{k}\left(
x\right) \left( \mathring{R}_{\text{ }ij\lambda }^{\alpha }+2\mathring{\nabla%
}_{[j}T_{\text{ }i]\lambda }^{\alpha }+2T_{\text{ }[j|\beta }^{\alpha
}T_{|i]\lambda }^{\;\;\;\;\beta }\right) \text{,}
\end{equation}%
where\textbf{\ }$\mathring{\nabla}_{\mu }A^{\alpha }:=\partial _{\mu
}A^{\alpha }+\mathring{\Gamma}_{\mu \beta }^{\alpha }A^{\beta }$, $\mathring{%
\nabla}_{\mu }A_{\alpha }:=\partial _{\mu }A_{\alpha }-\mathring{\Gamma}%
_{\mu \alpha }^{\beta }A_{\beta }$, the square brackets in $T_{\text{ }%
[j|\beta }^{\alpha }T_{|i]\lambda }^{\;\;\;\;\beta }$ represents
anti-symmetrization with respect to $ij$, $\beta $ being fixed and $%
\mathring{R}_{\text{ \ }\gamma \rho \lambda }^{\alpha }=R_{\text{ \ }\gamma
\rho \lambda }^{\alpha }\left( \Gamma \rightarrow \mathring{\Gamma}\right) $%
. We note that the Ricci tensor $R_{\mu \lambda }=R_{\mu \alpha \lambda
}^{\;\;\;\;\ \alpha }$ takes the form%
\begin{equation}
R_{\mu \lambda }=\mathring{R}_{\mu \lambda }\left( \mathring{\Gamma}\right) +%
\mathring{\nabla}_{\alpha }T_{\mu \lambda }^{\;\;\;\alpha }-\mathring{\nabla}%
_{\mu }T_{\alpha \lambda }^{\;\;\;\alpha }+T_{\alpha \beta }^{\;\;\;\alpha
}T_{\mu \lambda }^{\;\;\;\beta }-T_{\mu \beta }^{\;\;\;\alpha }T_{\alpha
\lambda }^{\;\;\;\beta }\text{,}
\end{equation}%
where the torsion-free contribution $\mathring{R}_{\mu \lambda }\left( 
\mathring{\Gamma}\right) $ is defined as, 
\begin{equation}
\mathring{R}_{\mu \nu }\left( \mathring{\Gamma}\right) =\partial _{\gamma }%
\mathring{\Gamma}_{\mu \nu }^{\gamma }-\partial _{\nu }\mathring{\Gamma}%
_{\mu \gamma }^{\gamma }+\mathring{\Gamma}_{\mu \nu }^{\gamma }\mathring{%
\Gamma}_{\gamma n}^{n}-\mathring{\Gamma}_{\mu k}^{\gamma }\mathring{\Gamma}%
_{\nu \gamma }^{k}\text{.}
\end{equation}%
From (\ref{inter6a}) we can deduce a relation that allows to compute the
affine connection in terms of the spin connection (and tetrad) or
vice-versa, namely \cite{crawford}%
\begin{equation}
\Gamma ^{\sigma }\,_{\mu \nu }=e^{a\sigma }\left( \partial _{\mu }e_{a\nu
}-\omega _{\text{ \ }a\mu }^{b}e_{b\nu }\right) \text{.}  \label{gamma_omega}
\end{equation}%
It is interesting to observe that substituting $\Gamma =\Gamma \left( \omega
\right) $ from (\ref{gamma_omega}) into (\ref{inter5}) leads to\textbf{\ }%
\begin{equation}
T_{\text{ }\beta \gamma }^{\alpha }e_{\alpha }^{i}e_{j}^{\beta
}e_{k}^{\gamma }=e_{j}^{\beta }e_{k}^{\gamma }\left( D_{\beta }e_{\gamma
}^{i}-D_{\gamma }e_{\beta }^{i}\right) \equiv C_{\text{ }jk}^{i}\text{,}
\label{torsion}
\end{equation}%
which establishes a means to transform between the holonomic torsion tensor%
\textbf{\ }$T_{\text{ }\beta \gamma }^{\alpha }$\textbf{\ }in (\ref{inter5})
and the non-holonomic structure functions\textbf{\ }$C_{\text{ }jk}^{i}$%
\textbf{\ }in (\ref{struct}) (and vice-versa) in terms of appropriate tetrad
index saturation. This situation is entirely analogous to the transformation
from $R_{\text{ \ }jkl}^{i}\left( \omega \right) $ to $R_{\text{ \ }\gamma
\rho \lambda }^{\alpha }\left( \Gamma \right) $ (and vice-versa) via tetrad
index saturation. From (\ref{torsion}) or (\ref{struct}), the torsion tensor
can be viewed as a sort of field strength associated with the tetrad
coefficients that describes a twist of the tetrad under parallel transport
(relative to a given basis) that is independent of the effect of curvature
(i.e., a twist in a plane perpendicular to the plane of parallel transport).
This is to be compared with the interpretation of torsion as the asymmetric
part of the affine connection according to (\ref{inter5}). Equation (\ref%
{torsion}) can be solved for the spin connection, yielding \cite{Blagojevic}%
\textbf{\ }%
\begin{equation}
\omega _{ab\mu }:=\frac{1}{2}\left( \Omega _{cab}+\Omega _{bca}-\Omega
_{abc}\right) e_{\text{ \ }\mu }^{c}\left( x\right) +T_{ab\mu }\text{.}
\end{equation}%
The quantities $T_{ab\mu }$ are related to the spacetime torsion tensor $%
T_{\alpha \beta \mu }$ according to $T_{ab\mu }:=e_{j}^{\text{ }\alpha
}\left( x\right) e_{k}^{\beta }\left( x\right) T_{\alpha \beta \mu }$. We
assume in this work that the torsion is totally antisymmetric and of
potential type, that is, we employ the \textit{ansatz} that $T_{\text{ \ }%
\nu \mu }^{\lambda }$ is derived from a second-rank, tensor potential $%
H_{\mu \nu }=-H_{\nu \mu }$ according to \cite{Hammond1} 
\begin{equation}
T_{\rho \beta \gamma }\overset{\text{def}}{=}\partial _{\lbrack \rho
}H_{\beta \gamma ]}\text{.}  \label{tor-def}
\end{equation}%
The Lagrangian density for a fermion field $\psi \left( x\right) $ in curved
spacetime \cite{Wald, BirrellDavies} with torsion is given by%
\begin{equation}
\mathcal{L}_{\text{matter}}=\frac{i}{2}\left[ \left( D_{\mu }\bar{\psi}%
\right) \gamma ^{\mu }\psi -\bar{\psi}\gamma ^{\mu }D_{\mu }\psi \right] -m%
\bar{\psi}\psi -eA_{\mu }j_{\left( e\right) }^{\mu }\text{, \ }j_{\left(
e\right) }^{\mu }:=i\bar{\psi}\gamma ^{\mu }\psi \text{,}  \label{MatLag}
\end{equation}%
where $\bar{\psi}$ is the Pauli conjugate of the Dirac field $\psi $ defined
by $\bar{\psi}(x)=i\psi ^{\dagger }(x)\,\gamma _{0}$, ($\dagger $) is the
Hermitian conjugate and $\gamma $ represents the appropriate Dirac $\gamma $%
-matrix with $\gamma ^{\mu }:=e_{i}^{\mu }(x)\gamma ^{i}$, $A_{\mu }$ is the
electromagnetic $4$-vector potential, $e$ is the electric charge of the
fermion and $j_{\left( e\right) }^{\mu }$ is the fermion current. The
Lagrangian density (\ref{MatLag}) can be re-written as%
\begin{equation}
\mathcal{L}_{\text{matter}}=\mathcal{\mathring{L}}_{\text{matter}}-\frac{1}{8%
}T_{\mu \alpha \beta }\bar{\psi}\left\{ \gamma ^{\mu }\text{, }\gamma
^{\alpha \beta }\right\} \psi -eA_{\mu }j_{\left( e\right) }^{\mu }\text{, \ 
}j_{\left( e\right) }^{\mu }:=i\bar{\psi}\gamma ^{\mu }\psi \text{,}
\end{equation}%
where%
\begin{equation}
\mathcal{\mathring{L}}_{\text{matter}}=\frac{i}{2}\left[ \left( \mathring{D}%
_{\alpha }\bar{\psi}\right) \gamma ^{\alpha }\psi -\bar{\psi}\gamma ^{\alpha
}\mathring{D}_{\alpha }\psi \right] -m\bar{\psi}\psi \text{,}
\end{equation}%
with $\mathring{D}_{\alpha }\psi :=\partial _{\alpha }\psi -\frac{1}{4}%
\mathring{\omega}_{\alpha ij}\gamma ^{ij}\psi $ and $\mathring{D}_{\alpha }%
\bar{\psi}:=\partial _{\alpha }\bar{\psi}+\frac{1}{4}\mathring{\omega}%
_{\alpha ij}\bar{\psi}\gamma ^{ij}$, $\mathring{\omega}_{\alpha ij}=\frac{1}{%
2}e_{\text{ \ }\alpha }^{c}\left( x\right) \left( \Omega _{cij}+\Omega
_{jci}-\Omega _{ijc}\right) $ being the torsion-free spin connection. Using
the following relations%
\begin{equation}
\left\{ 
\begin{array}{c}
-\frac{1}{4}T_{\mu \alpha \beta }\bar{\psi}\left\{ \gamma ^{\mu }\text{, }%
\gamma ^{\alpha \beta }\right\} \psi =\frac{1}{4}T_{\mu \alpha \beta }\bar{%
\psi}\left( \gamma ^{\beta \alpha }\gamma ^{\mu }-\gamma ^{\mu }\gamma
^{\alpha \beta }\right) \psi \text{,} \\ 
\\ 
\gamma ^{\mu }\gamma ^{\nu }\gamma ^{\lambda }\varepsilon _{\mu \nu \lambda
\sigma }=\left\{ \gamma ^{\mu }\text{, }\gamma ^{\nu \lambda }\right\}
\varepsilon _{\mu \nu \lambda \sigma }=3!\gamma _{\sigma }\gamma _{5}\text{, 
}\left\{ \gamma ^{\mu }\text{, }\gamma ^{\nu \lambda }\right\} =\gamma
^{\lbrack \mu }\gamma ^{\nu }\gamma ^{\lambda ]}\text{,}%
\end{array}%
\right. 
\end{equation}%
we obtain%
\begin{equation}
T_{\mu \alpha \beta }\bar{\psi}\left\{ \gamma ^{\mu }\text{, }\gamma
^{\alpha \beta }\right\} \psi =\frac{1}{2i}T_{\mu \alpha \beta }\varepsilon
^{\alpha \beta \mu \nu }j_{5\nu }\text{, \ }j_{5\nu }:=\bar{\psi}\gamma
_{5}\gamma _{\nu }\psi \text{,}  \label{inter}
\end{equation}%
where $j_{5\nu }$ is the fermion pseudo-current. Defining the torsion
axial-vector (also referred to as the torsion dual in what follows)%
\begin{equation}
T^{\nu }:=\frac{1}{3!}\varepsilon ^{\alpha \beta \mu \nu }T_{\alpha \beta
\mu }\text{.}
\end{equation}%
the first equation in (\ref{inter}) becomes, 
\begin{equation}
\left( \bar{\psi}\gamma _{5}\gamma _{\nu }\psi \right) \varepsilon ^{\alpha
\beta \mu \nu }T_{\mu \alpha \beta }=-6ij_{5\nu }T^{\nu }\text{.}
\end{equation}%
The interaction between the Dirac field and torsion has been reduced to a
coupling of the fermion axial current to a torsion axial-vector $T_{\mu }$.
Thus, the matter Lagrangian density in curved space with torsion \cite%
{Carroll} and electromagnetic fields reads%
\begin{equation}
\mathcal{L}_{\text{matter}}=\mathcal{\mathring{L}}_{\text{matter}}+\frac{3i}{%
8}T_{\mu }j_{5}^{\mu }-eA_{\mu }j_{\left( e\right) }^{\mu }\text{.}
\label{matter}
\end{equation}

\section{The Total System Lagrangian Density}

We now consider the geometrical setting in which the matter content -
represented by Lagrangian density (\ref{matter}) - is immersed. The
Einstein-Hilbert Lagrangian density is given by 
\begin{equation}
\mathcal{L}_{\text{geom}}=\sqrt{-g}\frac{R}{k_{0}}\text{,}
\end{equation}%
where $k_{0}=\frac{16\pi G}{c^{4}}$ and $R=g^{ij}R_{ij}$ is the scalar
curvature. Note that $l_{p}=\left( G\right) ^{1/2}$ is the Planck constant
(in geometrized units). In the Brans-Dicke generalization of gravity, one
introduces a scalar field $\Phi $ via the replacement $G\rightarrow e^{2\Phi
}G$ $\left( \text{i.e. }k_{0}\rightarrow e^{2\Phi }k_{0}\right) $. For
simplicity, let $\alpha =\frac{e^{-2\Phi }}{k_{0}}$. With this
generalization and the transformation\textbf{\ }$k_{0}\rightarrow e^{2\Phi
}k_{0}$, the geometrical Lagrangian density becomes 
\begin{equation}
\mathcal{L}_{\text{geom}}=\sqrt{-g}\alpha R=\sqrt{-g}\frac{e^{-2\Phi }}{k_{0}%
}\left( \mathring{R}+\partial _{\gamma }T_{\alpha }^{\text{ \ }\gamma \alpha
}+T_{\alpha }^{\;\;\beta \lambda }T_{\lambda \beta }^{\;\;\;\alpha }\right) 
\text{.}  \label{L-geom}
\end{equation}%
Observe that the quantity $\partial _{\gamma }T_{\alpha }^{\text{ \ }\gamma
\alpha }$ in (\ref{L-geom}) is vanishing since $T_{\alpha }^{\text{ \ }%
\gamma \alpha }=0$ (due to the total antisymmetry of the torsion tensor, see
(\ref{tor-def})). For this reason, we may choose to rewrite $e^{-2\Phi
}\partial _{\gamma }T_{\alpha }^{\text{ \ }\gamma \alpha }$ as a total
divergence $\partial _{\gamma }\left( e^{-2\Phi }T_{\alpha }^{\text{ \ }%
\gamma \alpha }\right) $\textbf{\ }which does not contribute to the
equations of motion\textbf{.} We work in the so-called Einstein frame \cite%
{Gasparini, cho}. For this reason we perform a conformal transformation on
the metric tensor%
\begin{equation}
g_{\mu \nu }\rightarrow g_{\mu \nu }^{\prime }=e^{2\Phi }g_{\mu \nu }\text{,
\ }g^{\mu \nu }\rightarrow g^{\prime \mu \nu }=e^{-2\Phi }g^{\mu \nu }\text{,%
}  \label{conformal}
\end{equation}%
which leads to%
\begin{equation}
\sqrt{-g^{\prime }}=e^{4\Phi }\sqrt{-g}\text{,}
\end{equation}%
where $g=\det g_{\mu \nu }$ and $\ g^{\prime }=e^{8\Phi }g$. Under the
conformal transformation (\ref{conformal}) the Christoffel symbols transform
according to,%
\begin{equation}
\mathring{\Gamma}_{\mu \nu }^{\alpha }\rightarrow \mathring{\Gamma}^{\prime
}{}_{\mu \nu }^{\alpha }=\mathring{\Gamma}_{\mu \nu }^{\alpha }+g^{\alpha
\beta }\left[ \left( \partial _{\mu }\Phi \right) g_{\nu \beta }+\left(
\partial _{\nu }\Phi \right) g_{\beta \mu }-\left( \partial _{\beta }\Phi
\right) g_{\mu \nu }\right] \text{.}  \label{Christoffel-transf}
\end{equation}%
With the conformally transformed Christoffel symbols (\ref%
{Christoffel-transf}), the correspondingly transformed scalar curvature is
given by%
\begin{equation}
\mathring{R}\rightarrow \mathring{R}^{\prime }=g^{\prime \mu \nu }\left(
\partial _{\nu }\mathring{\Gamma}^{\prime }\text{ }_{\beta \mu }^{\beta
}-\partial _{\beta }\mathring{\Gamma}^{\prime }{}_{\nu \mu }^{\beta }+%
\mathring{\Gamma}^{\prime }{}_{\nu \lambda }^{\beta }\mathring{\Gamma}%
^{\prime }{}_{\beta \mu }^{\lambda }-\mathring{\Gamma}^{\prime }{}_{\beta
\lambda }^{\beta }\mathring{\Gamma}^{\prime }{}_{\nu \mu }^{\lambda }\right) 
\text{.}
\end{equation}%
By direct calculation, we obtain%
\begin{equation}
\mathring{R}^{\prime }=e^{2\Phi }\mathring{R}+6e^{2\Phi }g^{\mu \nu }\left[
\left( \partial _{\mu }\Phi \right) \left( \partial _{\nu }\Phi \right) +%
\frac{1}{2}\partial _{\mu }\partial _{\nu }\Phi \right] \text{.}
\end{equation}%
Letting $\phi =2\Phi $, the geometrical Lagrangian density becomes%
\begin{equation}
\mathcal{L}_{\text{geom}}^{\prime }=\frac{1}{k_{0}}\left[ \mathring{R}%
-e^{-2\phi }T_{\mu \nu \sigma }T^{\mu \nu \sigma }+\frac{3}{2}\left(
\partial ^{\mu }\phi \right) \left( \partial _{\mu }\phi \right) +\frac{3}{2}%
\square \phi \right] \text{,}  \label{ECW}
\end{equation}%
where $\square :=g^{\mu \nu }\partial _{\mu }\partial _{\nu }$.\textbf{\ }It
is worth observing that the dilaton kinetic term in is generated by the
conformal transformation (\ref{conformal}) acting on the curvature scalar
taking $\mathring{R}$ to $\mathring{R}^{\prime }$. Moreover, we note that
the Lagrangian density (\ref{ECW})\ is true up to a total divergence that is
proportional to $\partial _{\gamma }\left( e^{-2\Phi }T_{\alpha }^{\text{ \ }%
\gamma \alpha }\right) $. As a working hypothesis we assume the dilaton%
\textbf{\ }$\phi $\textbf{\ }possess an isotropic field configuration (i.e.%
\textbf{\ }$\phi \left( \vec{r}\right) =\phi \left( \left\vert \vec{r}%
\right\vert \right) $).

It is straightforward to verify that under conformal transformation (\ref%
{conformal}) $\mathcal{L}_{\text{matter}}$ is invariant. Using the conformal
transformation on spinor fields \cite{Shapiro}%
\begin{equation}
\psi \rightarrow \psi ^{\prime }=e^{-\frac{3}{2}\phi \left( \left\vert \vec{r%
}\right\vert \right) }\psi \text{, \ }\bar{\psi}\rightarrow \bar{\psi}%
^{\prime }=e^{\frac{3}{2}\phi \left( \left\vert \vec{r}\right\vert \right) }%
\bar{\psi}\text{,}
\end{equation}%
we determine$\ $%
\begin{equation}
\mathcal{\mathring{L}}_{\text{matter}}\rightarrow \mathcal{\mathring{L}}_{%
\text{matter}}^{\prime }=\mathcal{\mathring{L}}_{\text{matter}}+3\partial
_{\mu }\left( \phi j_{\left( e\right) }^{\mu }\right) -3\phi \partial _{\mu
}j_{\left( e\right) }^{\mu }\text{.}
\end{equation}%
If the fermion current is conserved, then we expect $\partial _{\mu
}j_{\left( e\right) }^{\mu }=0$. Since $\partial _{\mu }\left( \phi
j_{\left( e\right) }^{\mu }\right) $ is a total divergence it does not
contribute to the equations of motion so it may be ignored. The interaction
term $\frac{3}{4}T_{\mu \nu \sigma }\bar{\psi}\gamma ^{\lbrack \mu }\gamma
^{\nu }\gamma ^{\sigma ]}\psi $ is invariant under the conformal
transformations (\ref{conformal}) since $T_{\mu \nu \sigma }$ is postulated
to be so, and the spin energy potential $\tau ^{\mu \nu \sigma }:=\bar{\psi}%
\gamma ^{\lbrack \mu }\gamma ^{\nu }\gamma ^{\sigma ]}\psi $ is trivially
invariant under scale transformations. It is obvious that the mass term is
invariant under scale transformations.

Having introduced the geometrical setting and matter content (electrically
charged, nonrelativistic spin-$\frac{1}{2}$ particles) of the model,\ we now
consider electromagnetic interaction between such prototype matter and
abelian magnetic monopoles, where the photons mediating this interaction are
topologically coupled to the anti-symmetric torsion potential. We are
concerned with investigating whether the incompatibility of massive photons
and magnetic monopoles within a classical theory is a consequence of the 
\textit{a priori} inclusion of photon mass or is related to the specific
mechanism for gauge boson mass generation. We include the monopole in a
non-dynamical manner.\textbf{\ }The Lagrangian density describing the gauge
sector of this scenario is given by%
\begin{equation}
\mathcal{L}_{\text{gauge}}=-\frac{1}{4}\mathcal{F}^{\mu \nu }\mathcal{F}%
_{\mu \nu }+\mu _{0}\varepsilon _{\alpha \beta \rho \sigma }A^{\alpha
}\partial ^{\beta }H^{\rho \sigma }\text{, }\mathcal{F}_{\mu \nu }:=F_{\mu
\nu }+\text{ }^{\ast }\mathcal{G}_{\mu \nu }\text{,}
\end{equation}%
where the Hodge dual $\left( \ast \right) $ of $\mathcal{G}^{\mu \nu }$ is
defined by $^{\ast }\mathcal{G}^{\mu \nu }:=\frac{1}{2!}\epsilon ^{\mu \nu
\rho \sigma }\mathcal{G}_{\rho \sigma }$. The electromagnetic field strength
has the usual form $F_{\nu \sigma }:=\partial _{\nu }A_{\sigma }-\partial
_{\sigma }A_{\nu }$ while the monopole contribution is given by 
\begin{equation}
\mathcal{G}^{\rho \sigma }\left( \vec{r}\right) :=4\pi e_{\left( m\right)
}\int dx^{\rho }\wedge dx^{\sigma }\delta ^{\left( 4\right) }\left( \vec{r}-%
\vec{r}_{\text{monopole}}\right) \text{,}  \label{mono}
\end{equation}%
where $e_{\left( m\right) }$ is the magnetic charge. Observe that (\ref{mono}%
) is antisymmetric and is responsible for breaking the Bianchi identity (\ref%
{A-BI}) in the\textbf{\ }$A_{\mu }$\textbf{\ }sector. The total Lagrangian
density $\mathcal{L}_{\text{total}}\left( \phi \text{, }H_{\mu \nu }\text{, }%
A_{\mu }\text{, }e_{\mu }^{i}\text{, }\psi \right) =\mathcal{L}_{\text{geom}%
}\left( \phi \text{, }e_{\mu }^{i}\right) +\mathcal{L}_{\text{gauge}}\left(
H_{\mu \nu }\text{, }A_{\mu }\right) +\mathcal{L}_{\text{matter}}\left( \psi 
\text{, }\bar{\psi}\right) $ is given by, 
\begin{eqnarray}
\mathcal{L}_{\text{total}} &=&\frac{1}{k_{0}}\left( \mathring{R}-e^{-2\phi
\left( \left\vert \vec{r}\right\vert \right) }T_{\mu \nu \sigma }T^{\mu \nu
\sigma }+\frac{3}{2}\left( \partial ^{\mu }\phi \right) \left( \partial
_{\mu }\phi \right) +\frac{3}{2}\square \phi \right) -\frac{1}{4}\mathcal{F}%
^{\mu \nu }\mathcal{F}_{\mu \nu }+\mu _{0}\varepsilon _{\alpha \beta \rho
\sigma }A^{\alpha }\partial ^{\beta }H^{\rho \sigma }+  \notag
\label{system} \\
&&  \notag \\
&&+\frac{i}{2}\left[ \left( \mathring{D}_{\mu }\bar{\psi}\right) \gamma
^{\mu }\psi -\bar{\psi}\gamma ^{\mu }\mathring{D}_{\mu }\psi \right] +\frac{%
3i}{8}T_{\mu }j_{5}^{\mu }-eA_{\mu }j_{\left( e\right) }^{\mu }-m\bar{\psi}%
\psi \text{.}
\end{eqnarray}%
We remark that the coupling term proportional to $\mu _{0}$ describes a
topological interaction between gauge fields $A^{\alpha }$ and $H^{\rho
\sigma }$. This may be understood from the lack of $\mu _{0}$-dependent
terms in the canonical energy-momentum tensor $\Sigma _{\mu \nu }$ appearing
in (\ref{energy-mom}). This fact reflects the lack of energy associated with
the interaction. Such interaction has no local propagating degrees of
freedom, hence being topological in nature \cite{Moura-Melo}.

\section{Field Equations}

By variation of the action $I=\int \sqrt{-g}d^{4}x\mathcal{L}_{\text{total}%
}\left( \phi \text{, }H_{\mu \nu }\text{, }A_{\mu }\text{, }e_{\mu }^{i}%
\text{, }\psi \right) $ with respect to $\phi $, $H_{\mu \nu }$, $A_{\mu }$
and $\bar{\psi}$, and requiring the coefficients of each variation
independently vanish, we obtain the equations of motion%
\begin{equation}
\left\{ 
\begin{array}{c}
\frac{\partial \mathcal{L}}{\partial \phi }-\partial _{\mu }\left( \frac{%
\partial \mathcal{L}}{\partial \left( \partial _{\mu }\phi \right) }\right)
=0\text{, }\frac{\partial \mathcal{L}}{\partial \bar{\psi}}-\partial _{\mu
}\left( \frac{\partial \mathcal{L}}{\partial \left( \partial _{\mu }\bar{\psi%
}\right) }\right) =0\text{,} \\ 
\\ 
\frac{\partial \mathcal{L}}{\partial H_{\mu \nu }}-\partial _{\sigma }\left( 
\frac{\partial \mathcal{L}}{\partial \left( \partial _{\sigma }H_{\mu \nu
}\right) }\right) =0\text{, }\frac{\partial \mathcal{L}}{\partial A_{\mu }}%
-\partial _{\sigma }\left( \frac{\partial \mathcal{L}}{\partial \left(
\partial _{\sigma }A_{\mu }\right) }\right) =0\text{.}%
\end{array}%
\right. 
\end{equation}%
To obtain the explicit form of the dynamical equations for the fermions we
recall that the Dirac $\gamma $-matrices are covariantly constant,%
\begin{equation}
\nabla _{\kappa }\gamma _{\iota }=\partial _{\kappa }\gamma _{\iota }-\Gamma
_{\iota \kappa }^{\mu }\gamma _{\mu }+\left[ \gamma _{\iota }\text{, }\hat{%
\Gamma}_{\kappa }\right] =0\text{ with }\hat{\Gamma}_{\kappa }=\frac{1}{8}%
\left[ \left( \partial _{\kappa }\gamma _{\iota }\right) \gamma ^{\iota
}-\Gamma _{\text{ \ }\iota \kappa }^{\mu }\gamma _{\mu }\gamma ^{\iota }%
\right] 
\end{equation}%
The $4\times 4$ matrices $\hat{\Gamma}_{\kappa }$ are real matrices used to
induce similarity transformations on quantities with spinor transformation
properties \cite{Brill}, that is $\gamma _{i}\rightarrow \gamma _{i}^{\prime
}=\hat{\Gamma}^{-1}\gamma _{i}\hat{\Gamma}$. Varying $\hat{\Gamma}_{\kappa }$
leads to $\delta \hat{\Gamma}_{\kappa }=\frac{1}{8}\left[ \left( \partial
_{\kappa }\delta \gamma _{\iota }\right) \gamma ^{\iota }-\left( \delta
\Gamma _{\text{ \ }\iota \kappa }^{\mu }\right) \gamma _{\mu }\gamma ^{\iota
}\right] $. Since we require the anticommutator condition on the gamma
matrices $\gamma _{\mu }\gamma _{\nu }+\gamma _{\nu }\gamma _{\mu }=g_{\mu
\nu }\mathbf{1}$ (Dirac algebra) to hold, the variation of the metric gives%
\begin{equation}
2\delta g^{\mu \nu }=\{\delta \gamma ^{\mu }\text{, }\gamma ^{\nu
}\}+\{\gamma ^{\mu }\text{, }\delta \gamma ^{\nu }\}\text{.}
\end{equation}%
One solution to this equation is $\delta \gamma ^{\nu }=\frac{1}{2}\gamma
_{\sigma }\delta \gamma ^{\sigma \nu }$. With the aid of this result, we can
write $\left( \partial _{\kappa }\delta \gamma _{\iota }\right) \gamma
^{\iota }=\frac{1}{2}\partial _{\kappa }\left( \gamma ^{\nu }\delta g_{\nu
\iota }\right) \gamma ^{\iota }$. Finally, exploiting the anti-symmetry in $%
\gamma _{\mu \nu }$ we obtain%
\begin{equation}
\delta \hat{\Gamma}_{\kappa }=\frac{1}{8}\left( g_{\nu \sigma }\delta \Gamma
_{\mu \kappa }^{\text{ \ \ }\sigma }-g_{\mu \sigma }\delta \Gamma _{\nu
\kappa }^{\text{ \ \ }\sigma }\right) \gamma ^{\mu \nu }\text{.}
\end{equation}%
With the above variational relations, it is straightforward to show that the
dynamical equation for the fermions is a nonlinear Dirac equation \cite%
{Hehl2} of Heisenberg-Pauli type,%
\begin{equation}
\left[ \gamma ^{\mu }\left( \mathring{D}_{\mu }-ieA_{\mu }\right) +\frac{3}{8%
}T_{\mu \nu \sigma }\gamma ^{\lbrack \mu }\gamma ^{\nu }\gamma ^{\sigma ]}-m%
\right] \psi =0\text{.}
\end{equation}%
For the scalar field $\phi $ we obtain the Klein-Gordon equation%
\begin{equation}
\square \phi \left( \left\vert \vec{r}\right\vert \right) -\frac{4}{3}%
e^{-2\phi \left( \left\vert \vec{r}\right\vert \right) }T_{\mu \nu \sigma
}T^{\mu \nu \sigma }=0\text{.}
\end{equation}%
To obtain the analogue of the Einstein equations the following calculations
involving the metric tensor $g_{\mu \nu }$ and its determinant $g=\det
\left( g_{\mu \nu }\right) $ are useful. Recall $gg^{\mu \nu }=\frac{%
\partial g}{\partial g_{\mu \nu }}$ and\ $gg_{\mu \nu }=-\frac{\partial g}{%
\partial g^{\mu \nu }}$. Now since $\delta \sqrt{-g}=\frac{\partial \sqrt{-g}%
}{\partial g}\delta g=-\frac{\delta g}{2\sqrt{-g}}$ where $\frac{\delta g}{%
\delta g_{\mu \nu }}=gg^{\mu \nu }$,\ we can write $\delta g=gg^{\mu \nu
}\delta g_{\mu \nu }$. Thus, we obtain $\delta \sqrt{-g}=-\frac{gg^{\mu \nu
}\delta g_{\mu \nu }}{2\sqrt{-g}}$. However, since $gg^{\mu \nu }\delta
g_{\mu \nu }=gg_{\mu \nu }\delta g^{\mu \nu }=\sqrt{-g}\sqrt{-g}g_{\mu \nu
}\delta g^{\mu \nu }$, we conclude $\frac{gg^{\mu \nu }\delta g_{\mu \nu }}{%
\sqrt{-g}}=\sqrt{-g}g_{\mu \nu }\delta g^{\mu \nu }$. Hence, 
\begin{equation}
\delta \sqrt{-g}=-\frac{1}{2}\sqrt{-g}g_{\mu \nu }\delta g^{\mu \nu }\text{.}
\end{equation}%
Writing the metric in terms of the tetrads $g^{\mu \nu }=e_{\;i}^{\mu
}e^{\nu i}$, we observe $\delta \sqrt{-g}=-\frac{1}{2}\sqrt{-g}\left( \delta
e_{\;i}^{\mu }e_{\mu }^{\;i}+e_{\nu i}\delta e^{\nu i}\right) $. By using $%
\delta e^{\nu i}=\delta \left( \eta ^{ij}e_{\;j}^{\nu }\right) =\eta
^{ij}\delta e_{\;j}^{\nu }$, we are able to deduce%
\begin{equation}
\delta \sqrt{-g}=-\sqrt{-g}e_{\mu }^{\;i}\delta e_{i}^{\;\mu }\text{.}
\end{equation}%
To compute the variation of the scalar curvature $R$ we must consider the
variation of the ordinary Ricci tensor $\mathring{R}_{i\nu }=e_{i}^{\;\mu }%
\mathring{R}_{\mu \nu }$ which is given by $\delta \mathring{R}_{i\nu
}=\delta e_{i}^{\;\mu }\mathring{R}_{\mu \nu }+e_{i}^{\;\mu }\delta 
\mathring{R}_{\mu \nu }$. In an inertial frame the Ricci tensor reduces to $%
\mathring{R}_{\mu \nu }=\partial _{\nu }\mathring{\Gamma}_{\beta \mu
}^{\beta }-\partial _{\beta }\mathring{\Gamma}_{\nu \mu }^{\beta }$ so that $%
\delta \mathring{R}_{i\nu }=\delta e_{i}^{\;\mu }\mathring{R}_{\mu \nu
}+e_{i}^{\;\mu }\left( \partial _{\nu }\delta \mathring{\Gamma}_{\beta \mu
}^{\beta }-\partial _{\beta }\delta \mathring{\Gamma}_{\nu \mu }^{\beta
}\right) $. The second term can be converted into a surface term and does
not contribute to the field equations, so it may be ignored leading to
conclude%
\begin{equation}
\delta \mathring{R}_{i\nu }=\delta e_{i}^{\;\mu }\mathring{R}_{\mu \nu }%
\text{.}
\end{equation}%
With the aid of $\delta \mathring{R}_{i\nu }$\ we may write the variation $%
\delta R$ as%
\begin{equation}
\delta R=\mathring{R}^{\mu \nu }\delta g_{\mu \nu }+g^{\mu \nu }\left(
\nabla _{\lambda }\delta \mathring{\Gamma}_{\text{ \ \ }\mu \nu }^{\lambda
}-\nabla _{\nu }\delta \mathring{\Gamma}\text{ }_{\text{ \ }\mu \lambda
}^{\lambda }\right) -T_{\alpha }^{\text{ \ }\beta \gamma }\delta T_{\beta
\gamma }^{\text{ \ \ }\alpha }\text{.}
\end{equation}%
With the above variational calculations involving the metric and Ricci
tensor it is not difficult to deduce the Einstein-like equations%
\begin{equation}
G_{\nu }^{\mu }+\Theta _{\nu }^{\mu }=k_{0}\Sigma _{\nu }^{\mu }\text{,}
\end{equation}%
with 
\begin{equation}
\Theta _{\nu }^{\mu }=-\left( 2T_{\nu \rho \sigma }T^{\mu \rho \sigma }+%
\mathring{\nabla}_{\sigma }T_{\nu }^{\;\mu \sigma }+\frac{3}{8}P_{\nu }^{\mu
}+e^{-\phi \left( \left\vert \vec{r}\right\vert \right) }Q_{\nu }^{\mu
}+\left( 1-e^{-2\phi \left( \left\vert \vec{r}\right\vert \right) }\right)
S_{\nu }^{\mu }\right) \text{,}
\end{equation}%
where%
\begin{equation}
\left\{ 
\begin{array}{c}
P_{\nu }^{\mu }=\delta _{\nu }^{\mu }\left( \partial ^{\sigma }\phi \right)
\left( \partial _{\sigma }\phi \right) -\left( \partial ^{\mu }\phi \right)
\left( \partial _{\nu }\phi \right) +\left( \delta _{\nu }^{\mu }\square
\phi -\partial ^{\mu }\partial _{\nu }\phi \right) \text{,} \\ 
\\ 
S_{\nu }^{\mu }=\delta _{\nu }^{\mu }T_{\lambda \alpha \sigma }T^{\lambda
\alpha \sigma }-3T_{\nu \alpha \sigma }T^{\mu \alpha \sigma }\text{, }Q_{\nu
}^{\mu }=\delta _{\nu }^{\mu }F_{\lambda \alpha }F^{\lambda \alpha }-F^{\mu
\alpha }F_{\alpha \nu }\text{.}%
\end{array}%
\right. 
\end{equation}%
The Einstein tensor $G_{\nu }^{\mu }$ is given by the standard form $G_{\nu
}^{\mu }=R_{\nu }^{\mu }-\frac{1}{2}R\delta _{\nu }^{\mu }$, while the
energy momentum tensor $\Sigma _{\mu \nu }$ reads%
\begin{equation}
\Sigma _{\mu \nu }=\bar{\psi}\gamma _{(\mu }\mathring{D}_{\nu )}\psi -%
\mathring{D}_{(\mu }\bar{\psi}\gamma _{\nu )}\psi +A_{(\mu }j_{\nu
)}^{\left( e\right) }-A_{\gamma }j_{(e)}^{\gamma }g_{\mu \nu }+T_{(\mu
}j_{5\nu )}-T_{\alpha }j_{5}^{\alpha }g_{\mu \nu }\text{.}
\label{energy-mom}
\end{equation}%
\ The Bianchi identities for the $A_{\mu }$ and $H_{\mu \nu }$-sectors read%
\begin{equation}
\partial _{\mu }\text{ }^{\ast }F^{\mu \nu }=-\partial _{\mu }\mathcal{G}%
^{\mu \nu }=-j_{\left( m\right) }^{\nu }\text{ and }\mathring{\nabla}%
_{\sigma }\left( e^{-2\phi \left( \left\vert \vec{r}\right\vert \right)
}T^{\sigma }\right) =0\text{,}  \label{A-BI}
\end{equation}%
where $j_{\left( m\right) }^{\nu }\equiv \left( \rho _{\left( m\right) }%
\text{, }\vec{j}_{\left( m\right) }\right) $, $\rho _{\left( m\right)
}=e_{\left( m\right) }\delta ^{\left( 3\right) }\left( \vec{r}-\vec{r}_{%
\text{monopole}}\right) $ and $\vec{j}_{\left( m\right) }=0$ since we work
in the monopole rest frame. For the gauge fields $A_{\beta }$\ and $H_{\rho
\sigma }$ we obtain the equations of motion%
\begin{equation}
\partial _{\mu }F^{\mu \nu }+2\mu _{0}T^{\nu }=ej_{\left( e\right) }^{\nu }%
\text{,}  \label{Cov-Max}
\end{equation}%
\begin{equation}
\frac{1}{k_{0}}\mathring{\nabla}_{\sigma }\left( e^{-2\phi \left( \left\vert 
\vec{r}\right\vert \right) }T^{\mu \nu \sigma }\right) -\mu _{0}\text{ }%
^{\ast }F^{\mu \nu }=\epsilon ^{\mu \nu \sigma \rho }\mathring{\nabla}%
_{\sigma }j_{\rho }^{5}\text{.}  \label{TorField}
\end{equation}%
By integrating (\ref{TorField}) we find the solution%
\begin{equation}
T^{\mu \nu \sigma }=e^{2\phi \left( \left\vert \vec{r}\right\vert \right) }%
\left[ k_{0}\epsilon ^{\mu \nu \sigma \rho }\left( j_{\rho }^{5}+\mu
_{0}A_{\rho }\right) +\Lambda ^{\mu \nu \sigma }\right] \text{,}
\label{TorsionSoln}
\end{equation}%
where $\Lambda ^{\mu \nu \sigma }$ arises from the process of integration
and must satisfy $\mathring{\nabla}_{\sigma }\Lambda ^{\mu \nu \sigma }=0$
and $\Lambda ^{(\mu \nu \sigma )}=0$. The general form of $\Lambda ^{\mu \nu
\sigma }$ that fulfills both conditions is $\Lambda ^{\mu \nu \sigma
}=\varepsilon ^{\mu \nu \sigma \gamma }\partial _{\gamma }f$, where $f$ is a
scalar function. From the solutions (\ref{TorsionSoln}) it is clear that the
sources of torsion are spinors, dilatons and the electromagnetic gauge
fields. Furthermore, equations (\ref{Cov-Max}), (\ref{TorField}) and (\ref%
{TorsionSoln}) describe a system of interacting charged fermions and abelian
magnetic monopoles where the interaction is mediated by topologically
massive vector boson with mass $m_{\mathcal{E}}^{2}=2\mu _{0}$.

\section{Generalized Maxwell's Equations, Classical Hamiltonian Formulation
and Magnetic Field Symmetry}

It is known that the classical non-relativistic theory describing the
massless electromagnetic scattering of an electric charge from a fixed
magnetic monopole has a well defined Hamiltonian formulation \cite{goldhaber}%
. Alternatively however, it is equally well known that one cannot construct
a self-consistent quantum field theory describing the nonrelativistic
electromagnetic interaction mediated by massive photons between a point-like
electric charge and a magnetic monopole \cite{joshi}. In one of our previous
work \cite{cafali}, we showed that this inconsistency arises in the
classical theory itself.

In this Section of the paper, we explore the possibility of constructing a
self-consistent nonrelativistic classical theory where magnetic monopoles
and topologically massive vector bosons coexist in the framework of
scalar-tensor gravity with torsion potential.

\subsection{Generalized Maxwell's Equations}

In this subsection, we begin by\ decomposing, for convenience, the
electromagnetic field strength and torsion tensors into their boost and
spatial components according to%
\begin{equation}
F_{\mu \nu }=\left\{ 
\begin{array}{c}
F_{0i}\equiv (\vec{E})_{i}\text{,} \\ 
F_{ij}\equiv -\epsilon _{ijk}(\vec{B})_{k}\text{,}%
\end{array}%
\right. \text{, }T_{\mu \nu \rho }=\left\{ 
\begin{array}{c}
T_{0ij}\equiv -\epsilon _{ijk}(\mathcal{\vec{E}})_{k}\text{,} \\ 
T_{ijk}\equiv \epsilon _{ijk}\mathcal{B}\text{,}%
\end{array}%
\right. \text{ and }T^{\mu }=\left( \mathcal{B}\text{,\ }\mathcal{\vec{E}}%
\right) 
\end{equation}%
and using the Bianchi identities (\ref{A-BI}) and the field equations (\ref%
{Cov-Max}) and (\ref{TorField}) we obtain the Maxwell-torsion equations in
standard vector notation%
\begin{equation}
\vec{\partial}\cdot \vec{E}\left( \vec{r}\right) =\rho _{\left( e\right)
}-2\mu _{0}\mathcal{B}\left( \vec{r}\right) \text{, }\mathring{\nabla}\cdot %
\left[ e^{-2\phi \left( \left\vert \vec{r}\right\vert \right) }\mathcal{\vec{%
E}}\left( \vec{r}\right) \right] =0\text{, }\vec{\partial}\cdot \vec{B}%
\left( \vec{r}\right) =\rho _{\left( m\right) }\text{,}  \label{mt1}
\end{equation}%
\begin{equation}
\vec{\partial}\times \vec{E}\left( \vec{r}\right) =-\partial _{t}\vec{B}%
\left( \vec{r}\right) \text{, }\mathring{\nabla}\times \left[ e^{-2\phi
\left( \left\vert \vec{r}\right\vert \right) }\mathcal{\vec{E}}\left( \vec{r}%
\right) \right] =k_{0}\mu _{0}\vec{B}\left( \vec{r}\right) +k_{0}\mathring{%
\nabla}\times \vec{j}_{5}\text{,}  \label{mt21}
\end{equation}%
\begin{equation}
\vec{\partial}\times \vec{B}\left( \vec{r}\right) =\vec{j}_{\left( e\right)
}+\partial _{t}\vec{E}\left( \vec{r}\right) -2\mu _{0}\mathcal{\vec{E}}%
\left( \vec{r}\right) \text{, }\mathring{\nabla}\left[ e^{-2\phi \left(
\left\vert \vec{r}\right\vert \right) }\mathcal{B}\left( \vec{r}\right) %
\right] =k_{0}\mathring{\nabla}\rho _{5}-k_{0}\mu _{0}\vec{E}\left( \vec{r}%
\right) \text{.}  \label{mt3}
\end{equation}%
where $\vec{\partial}$\ represents the ordinary nabla differential operator
of flat space, $j^{\mu \nu }=\mathring{\nabla}_{\sigma }\tau ^{\mu \nu
\sigma }=\epsilon ^{\mu \nu \alpha \beta }\mathring{\nabla}_{\alpha
}j_{\beta }^{5}$ and $j_{5}^{\beta }=\left( \rho _{5}\text{, }\vec{j}%
_{5}\right) $. The pseudo-current $j_{5}^{\beta }$\ arising from the spin
energy potential contributes to the diffusive magnetic potential. Observe
that the magnetic current $\vec{j}_{\mathbf{m}}$\ is absent from the first
equation in (\ref{mt21}) since we are working in the rest frame of the
monopole.\ In absence of electric fields, charges and currents, as well as
the absence of magnetic current and the zeroth component\textbf{\ }$\mathcal{%
B}$\textbf{\ }of the torsion dual $T^{\mu }$, the maxwell-torsion equations
become:%
\begin{equation}
\vec{\partial}\cdot \vec{E}\left( \vec{r}\right) =0\text{, }\mathring{\nabla}%
\cdot \left[ e^{-2\phi \left( \left\vert \vec{r}\right\vert \right) }%
\mathcal{\vec{E}}\left( \vec{r}\right) \right] =0\text{, }\vec{\partial}%
\cdot \vec{B}\left( \vec{r}\right) =\rho _{\left( m\right) }\text{,}
\end{equation}%
\begin{equation}
\vec{\partial}\times \vec{E}\left( \vec{r}\right) =0\text{, }\mathring{\nabla%
}\times \left[ e^{-2\phi \left( \left\vert \vec{r}\right\vert \right) }%
\mathcal{\vec{E}}\left( \vec{r}\right) \right] =k_{0}\mu _{0}\vec{B}\left( 
\vec{r}\right) +k_{0}\mathring{\nabla}\times \vec{j}_{5}\text{,}
\end{equation}%
\begin{equation}
\vec{\partial}\times \vec{B}\left( \vec{r}\right) =-2\mu _{0}\mathcal{\vec{E}%
}\left( \vec{r}\right) \text{.}
\end{equation}%
The total static magnetic field of the system is comprised of the point-like
magnetic charge, string, diffuse magnetic field (arising from the spatial
components\textbf{\ }$\mathcal{\vec{E}}$ of the torsion dual $T^{\mu }$) and
spin-magnetic (arising from $\vec{j}_{5}$) contributions%
\begin{eqnarray}
\vec{B}\left( \vec{r}\right)  &=&\vec{B}_{\text{monopole}}\left( \vec{r}%
\right) +\vec{B}^{\prime }\left( \vec{r}\right) =\left[ \vec{\partial}\times 
\vec{A}_{\text{monopole}}+e_{(m)}\vec{h}\left( \vec{r}\right) \right] +\vec{%
\partial}\times \left[ e^{-2\phi \left( \left\vert \vec{r}\right\vert
\right) }\mathcal{\vec{E}}\left( \vec{r}\right) +k_{0}\vec{j}_{5}\right]  
\notag \\
&=&\vec{\partial}\times \vec{A}+e_{(m)}\vec{h}\left( \vec{r}\right) \text{,}
\label{btotal}
\end{eqnarray}%
\textbf{\ }where $\vec{A}=\vec{A}_{\text{monopole}}+\vec{A}^{\prime }$\ with 
$\vec{A}^{\prime }=e^{-2\phi \left( \left\vert \vec{r}\right\vert \right) }%
\vec{E}+k_{0}\vec{j}_{5}$\ and the vector $\vec{A}_{\text{monopole}}$\ is a
singular vector potential representing the field of the fixed monopole%
\begin{equation}
\vec{A}_{\text{monopole}}(\vec{r})=\frac{e_{\left( m\right) }}{r^{2}}\frac{%
\sin (\theta )}{1+\cos (\theta )}(\hat{n}\times \vec{r})\text{, }\theta \neq
\pi \text{,}
\end{equation}%
with semi-infinite singularity line oriented along the negative $z$-axis.
The quantity $\vec{h}\left( \vec{r}\right) $\ is the magnetic string function%
\begin{equation}
\left\vert \vec{h}\left( \vec{r}\right) \right\vert =\frac{4\pi }{r^{2}}%
\frac{\delta \left( \theta \right) \delta \left( \varphi \right) }{\sin
\theta }\Theta \left( -\cos \theta \right) \text{, }\Theta \ \text{is the
Heaviside step function}
\end{equation}%
\textbf{\ }encountered in monopole theory. The magnetic field $\vec{B}_{%
\text{monopole}}$ in (\ref{btotal}) generated by the point-like magnetic
charge is given by%
\begin{equation}
\vec{B}_{\text{monopole}}\left( \vec{r}\right) =\frac{e_{\left( m\right) }%
\vec{r}}{r^{3}}\text{,}  \label{mag-mon}
\end{equation}%
whereas $\vec{B}^{\prime }\left( \vec{r}\right) $ in (\ref{btotal}) has the
form \cite{joshi, cafali} 
\begin{equation}
\vec{B}^{\prime }\left( \vec{r}\right) =b^{(1)}(r\text{, }\hat{n}\cdot \vec{r%
})\vec{r}+b^{(2)}(r\text{, }\hat{n}\cdot \vec{r})\hat{n}\text{,}
\label{bprime}
\end{equation}%
with $b^{(1)}$\ and $b^{(2)}$\ being general scalar field functions and $%
\widehat{n}$\ denoting a unitary vector along the monopole string. Combining
equations (\ref{mag-mon}) and (\ref{bprime}), equation (\ref{btotal}) becomes%
\begin{equation}
\vec{B}\left( \vec{r}\right) =\frac{e_{\left( m\right) }\vec{r}}{r^{3}}%
+b^{(1)}(r\text{, }\hat{n}\cdot \vec{r})\vec{r}+b^{(2)}(r\text{, }\hat{n}%
\cdot \vec{r})\hat{n}\text{.}  \label{babove}
\end{equation}%
It is clear from equation (\ref{babove}) that no spherically magnetic
solutions are allowed within Maxwell's generalized equations. Moreover, the
magnetic fields $\vec{B}_{\text{monopole}}\left( \vec{r}\right) $\ and $\vec{%
B}^{\prime }(\vec{r})$\ satisfy%
\begin{equation}
\vec{\partial}\cdot \vec{B}_{\text{monopole}}\left( \vec{r}\right)
=e_{\left( m\right) }\delta ^{(3)}(\vec{r})\text{\textbf{,}}\mathbf{\ }\vec{%
\partial}\times \vec{B}_{\text{monopole}}\left( \vec{r}\right) =0
\end{equation}%
and%
\begin{equation}
\vec{\partial}\cdot \vec{B}^{\prime }\left( \vec{r}\right) =0\text{, }\vec{%
\partial}\times \vec{B}^{\prime }\left( \vec{r}\right) =-m_{\mathcal{E}}^{2}(%
\vec{A}_{\text{monopole}}+e^{-2\phi \left( \left\vert \vec{r}\right\vert
\right) }\mathcal{\vec{E}})\text{, }m_{\mathcal{E}}^{2}=2\mu _{0}
\label{temp}
\end{equation}%
respectively. Notice that because of the second equation in (\ref{temp}) is
consistent with the non spherical symmetry of the total magnetic field in
equation (\ref{babove}).\ 

\subsection{Classical Hamiltonian Formulation and Magnetic Field Symmetry}

In this subsection, given the magnetic field solutions obtained in the
previous subsection, we consider the possibility of constructing a classical
non-relativistic Hamiltonian formulation of a theory describing a point-like
electric particle with charge $e$\ and mass $m$\ moving in the field of a
fixed monopole of charge $e_{\left( m\right) \text{ }}$. We require that the
constructed Poisson algebra of such Hamiltonian formulation be consistent
with the symmetry of the total magnetic vector field obtained in the above
subsection.

The Hamiltonian that describes to the above system is given by,\textbf{\ }%
\begin{equation}
H_{\text{total}}\overset{\text{def}}{=}\frac{(\mathring{\nabla}-e\vec{A})^{2}%
}{2m}+H_{\text{string}}\text{, }H_{\text{string}}=-ee_{\left( m\right) }\int
\left( \frac{d\vec{r}}{dt}\times \vec{h}\left( \vec{r}\right) \right) \cdot d%
\vec{r}\text{.}  \label{H}
\end{equation}%
The classical equation of motion arising from (\ref{H}), becomes%
\begin{equation}
m\frac{\mathring{\nabla}}{\mathring{\nabla}t}\vec{v}-e\frac{d\vec{r}}{dt}%
\times \left( \vec{\partial}\times \vec{A}\right) -ee_{\left( m\right) }%
\frac{d\vec{r}}{dt}\times \vec{h}\left( \vec{r}\right) =0\text{, }\vec{v}=%
\frac{d\vec{r}}{dt}\text{, }\vec{A}=\vec{A}_{\text{monopole}}+e^{-2\phi
\left( \left\vert \vec{r}\right\vert \right) }\mathcal{\vec{E}}
\end{equation}%
where in component form$\frac{\mathring{\nabla}v^{k}}{\mathring{\nabla}t}%
\overset{\text{def}}{=}\frac{dv^{k}}{dt}+\mathring{\Gamma}_{ij}^{k}v^{i}v^{j}
$ with $v^{k}=\frac{dr^{k}}{dt}$, $k=1$, $2$, $3$. For simplicity it is
assumed that $\left\vert \vec{j}_{5}\right\vert \ll \left\vert \frac{1}{k_{0}%
}e^{-2\phi \left( \left\vert \vec{r}\right\vert \right) }\mathcal{\vec{E}}%
\left( \vec{r}\right) \right\vert $ so we may neglect $\vec{\partial}\times 
\vec{j}_{5}$\ in the following analysis. Under this hypothesis, the total
magnetic field $\vec{B}\left( \vec{r}\right) $\ reduces to%
\begin{equation}
\vec{B}\left( \vec{r}\right) =\left[ \vec{\partial}\times \vec{A}_{\text{%
monopole}}+e_{(m)}\vec{h}\left( \vec{r}\right) \right] +\vec{\partial}\times %
\left[ e^{-2\phi \left( \left\vert \vec{r}\right\vert \right) }\mathcal{\vec{%
E}}\left( \vec{r}\right) \right] =\vec{\partial}\times \vec{A}+e_{(m)}\vec{h}%
\left( \vec{r}\right) \text{.}  \label{breduced}
\end{equation}%
Since we require that $\vec{B}\left( \vec{r}\right) $ be a vector field, we
must verify that the quantity $e^{-2\phi \left( \left\vert \vec{r}%
\right\vert \right) }\mathcal{\vec{E}}$\textbf{\ }transforms appropriately
under spatial rotations. Given that we are in a curved space, we must define
the spatial rotation generator associated to the Hamiltonian (\ref{H}) such
that it satisfies a proper Poisson algebra. We define the generator of
spatial rotations as $\vec{J}\overset{\text{def}}{=}\vec{L}+\vec{s}$\ such
that $\vec{J}\cdot \vec{s}=0$\ where $\vec{L}\overset{\text{def}}{=}\vec{r}%
\times \vec{P}$\ is the orbital angular momentum operator in curved space, $%
\vec{P}$ $\overset{\text{def}}{=}$ $\vec{p}-e\vec{A}$ is the curved space
kinetic momentum vector, $\vec{p}$ $\overset{\text{def}}{=}$\ $\vec{p}^{%
\text{flat}}-\mathring{\Gamma}$ is the curved space canonical momentum with $%
\vec{p}^{\text{flat}}\overset{\text{def}}{=}m\frac{d\vec{r}}{dt}+e\vec{A}$
being the ordinary canonical momentum vector of flat space. Finally $\vec{s}$
\ is defined as \cite{cafali}, 
\begin{equation}
\vec{s}\overset{\text{def}}{=}\int \left[ \vec{r}\times \left( \vec{E}\times 
\vec{B}\right) \right] d^{3}\vec{r}=\vec{s}_{\text{massless}}+e\int d\vec{r}%
\vec{r}\times \left[ \frac{\vec{r}}{r^{3}}\times \vec{B}^{\prime }\left( 
\vec{r}-\vec{R}\right) \right] 
\end{equation}%
with $\vec{s}_{\text{massless}}=ee_{\left( m\right) }\hat{R}$\ \cite%
{goldhaber, Berard} and $\vec{R}$\ is the relative vector position between
the monopole and the electric charge. The vector $\vec{s}$\ is taken as an
angular momentum with independent degrees of freedom and must obey the
following classical Poisson bracket relation%
\begin{equation}
\left\{ s_{i}\text{, }s_{j}\right\} =-\varepsilon _{ijk}s_{k}\text{.}
\end{equation}%
We make use of a result proved in \cite{Cap}, namely that in a curved
spacetime the fundamental Poisson brackets are always conserved. Thus, in
the curved spacetime that we consider, the Poisson brackets between two
generic functions $u(\vec{p}$, $\vec{r}$, $t)$\ and $g(\vec{p}$, $\vec{r}$, $%
t)$\ of the dynamical variables $\vec{p}$\ and $\vec{r}$, are defined in
usual manner as%
\begin{equation}
\left\{ u(\vec{p}\text{, }\vec{r}\text{, }t)\text{, }g(\vec{p}\text{, }\vec{r%
}\text{, }t)\right\} \overset{\text{def}}{=}\underset{i}{\sum }(\partial
_{p_{i}}u\partial _{r_{i}}g-\partial _{r_{i}}u\partial _{p_{i}}g)\text{.}
\end{equation}%
In what follows, we employ the basic canonical Poisson bracket structure for
the conjugate variables,%
\begin{equation}
\left\{ r_{i}\text{, }r_{j}\right\} =0\text{, }\left\{ r_{i}\text{, }p_{j}^{%
\text{flat}}\right\} =-\delta _{ij}\text{, }\left\{ p_{i}^{\text{flat}}\text{%
, }p_{j}^{\text{flat}}\right\} =0\text{.}
\end{equation}%
Since we are working within a curved spacetime geometry, it is necessary to
verify that the $\vec{J}$\ operators are in fact the generators of
rotations. To this end, we consider the Poisson bracket of $\vec{J}$\
operators,%
\begin{eqnarray}
\left\{ J_{i}\text{, }J_{l}\right\}  &=&\left\{ \varepsilon
_{ijk}r_{j}\left( p_{k}-A_{k}\right) +s_{i}\text{, }\varepsilon
_{lmn}r_{m}\left( p_{n}-A_{n}\right) +s_{l}\right\}   \label{poi3a} \\
&=&\left\{ \varepsilon _{ijk}r_{j}p_{k}-\varepsilon _{ijk}r_{j}A_{k}+s_{i}%
\text{, }\varepsilon _{lmn}r_{m}p_{n}-\varepsilon
_{lmn}r_{m}A_{n}+s_{l}\right\}   \notag \\
&=&\left\{ \varepsilon _{ijk}r_{j}p_{k}\text{, }\varepsilon
_{lmn}r_{m}p_{n}\right\} -\left\{ \varepsilon _{ijk}r_{j}p_{k}\text{, }%
\varepsilon _{lmn}r_{m}A_{n}\right\} +  \notag \\
&&-\left\{ \varepsilon _{ijk}r_{j}A_{k}\text{, }\varepsilon
_{lmn}r_{m}p_{n}\right\} +\left\{ \varepsilon _{ijk}r_{j}A_{k}\text{, }%
\varepsilon _{lmn}r_{m}A_{n}\right\} +\left\{ s_{i}\text{, }s_{l}\right\} 
\text{.}  \notag
\end{eqnarray}%
Note that in the previous Section and for the remainder of this subsection,
we set the electric charge $e=1$\ for convenience. The first bracket on the
right hand side (rhs) of (\ref{poi3a}) becomes%
\begin{eqnarray}
\left\{ \varepsilon _{ijk}r_{j}p_{k}\text{, }\varepsilon
_{lmn}r_{m}p_{n}\right\}  &=&\left\{ \varepsilon _{ijk}r_{j}\left( p_{k}^{%
\text{flat}}-\mathring{\Gamma}_{k}\right) \text{, }\varepsilon
_{lmn}r_{m}\left( p_{n}^{\text{flat}}-\mathring{\Gamma}_{n}\right) \right\} 
\\
&=&\left\{ \varepsilon _{ijk}r_{j}p_{k}^{\text{flat}}\text{, }\varepsilon
_{lmn}r_{m}p_{n}^{\text{flat}}\right\} -\left\{ \varepsilon
_{ijk}r_{j}p_{k}^{\text{flat}}\text{, }\varepsilon _{lmn}r_{m}\mathring{%
\Gamma}_{n}\right\} +  \notag \\
&&-\left\{ \varepsilon _{ijk}r_{j}\mathring{\Gamma}_{k}\text{, }\varepsilon
_{lmn}r_{m}p_{n}^{\text{flat}}\right\} +\left\{ \varepsilon _{ijk}r_{j}%
\mathring{\Gamma}_{k}\text{, }\varepsilon _{lmn}r_{m}\mathring{\Gamma}%
_{n}\right\} \text{,}  \notag
\end{eqnarray}%
where 
\begin{eqnarray}
\left\{ \varepsilon _{ijk}r_{j}p_{k}^{\text{flat}}\text{, }\varepsilon
_{lmn}r_{m}p_{n}^{\text{flat}}\right\}  &=&r_{l}p_{i}^{\text{flat}%
}-r_{i}p_{l}^{\text{flat}}\text{,} \\
-\left\{ \varepsilon _{ijk}r_{j}p_{k}^{\text{flat}}\text{, }\varepsilon
_{lmn}r_{m}\mathring{\Gamma}_{n}\right\}  &=&\delta _{il}r_{n}\mathring{%
\Gamma}_{n}-r_{l}\mathring{\Gamma}_{i}+\varepsilon _{ijk}\varepsilon
_{lmn}r_{m}p_{k}^{\text{flat}}\left\{ \mathring{\Gamma}_{n}\text{, }%
r_{j}\right\} \text{,} \\
-\left\{ \varepsilon _{ijk}r_{j}\mathring{\Gamma}_{k}\text{, }\varepsilon
_{lmn}r_{m}p_{n}^{\text{flat}}\right\}  &=&-\delta _{il}r_{k}\mathring{\Gamma%
}_{k}+r_{i}\mathring{\Gamma}_{l}+\varepsilon _{ijk}\varepsilon
_{lmn}r_{j}p_{n}^{\text{flat}}\left\{ r_{m}\text{, }\mathring{\Gamma}%
_{k}\right\} \text{,} \\
\left\{ \varepsilon _{ijk}r_{j}\mathring{\Gamma}_{k}\text{, }\varepsilon
_{lmn}r_{m}\mathring{\Gamma}_{n}\right\}  &=&-\varepsilon _{ijk}\varepsilon
_{lmn}r_{j}\mathring{\Gamma}_{n}\left\{ r_{m}\text{, }\mathring{\Gamma}%
_{k}\right\} -\varepsilon _{ijk}\varepsilon _{lmn}r_{m}\mathring{\Gamma}%
_{k}\left\{ \mathring{\Gamma}_{n}\text{, }r_{j}\right\} \text{.}
\end{eqnarray}%
Similarly, the second bracket on the rhs of (\ref{poi3a}) reduces to%
\begin{eqnarray}
-\left\{ \varepsilon _{ijk}r_{j}p_{k}\text{, }\varepsilon
_{lmn}r_{m}A_{n}\right\}  &=&-\left\{ \varepsilon _{ijk}r_{j}\left( p_{k}^{%
\text{flat}}-\mathring{\Gamma}_{k}\right) \text{, }\varepsilon
_{lmn}r_{m}A_{n}\right\}  \\
&=&-\left\{ \varepsilon _{ijk}r_{j}p_{k}^{\text{flat}}\text{, }\varepsilon
_{lmn}r_{m}A_{n}\right\} +\left\{ \varepsilon _{ijk}r_{j}\mathring{\Gamma}%
_{k}\text{, }\varepsilon _{lmn}r_{m}A_{n}\right\}   \notag
\end{eqnarray}%
where%
\begin{eqnarray}
-\left\{ \varepsilon _{ijk}r_{j}p_{k}^{\text{flat}}\text{, }\varepsilon
_{lmn}r_{m}A_{n}\right\}  &=&\delta _{il}r_{n}A_{n}-r_{l}A_{i}+\varepsilon
_{ijk}\varepsilon _{lmn}r_{m}p_{k}^{\text{flat}}\left\{ A_{n}\text{, }%
r_{j}\right\} \text{,} \\
+\left\{ \varepsilon _{ijk}r_{j}\mathring{\Gamma}_{k}\text{, }\varepsilon
_{lmn}r_{m}A_{n}\right\}  &=&-\varepsilon _{ijk}\varepsilon
_{lmn}r_{j}A_{n}\left\{ r_{m}\text{, }\mathring{\Gamma}_{k}\right\}
-\varepsilon _{ijk}\varepsilon _{lmn}r_{m}\mathring{\Gamma}_{k}\left\{ A_{n}%
\text{, }r_{j}\right\} \text{.}
\end{eqnarray}%
The third bracket on the rhs of (\ref{poi3a}) is similar to the second with $%
\vec{A}$\ and $\vec{p}$\ being interchanged such that,%
\begin{eqnarray}
-\left\{ \varepsilon _{ijk}r_{j}A_{k}\text{, }\varepsilon
_{lmn}r_{m}p_{n}\right\}  &=&-\delta _{il}r_{n}A_{n}+r_{i}A_{l}-\varepsilon
_{ijk}\varepsilon _{lmn}r_{j}p_{n}^{\text{flat}}\left\{ r_{m}\text{, }%
A_{k}\right\} + \\
&&+\varepsilon _{ijk}\varepsilon _{lmn}r_{m}A_{k}\left\{ \mathring{\Gamma}%
_{j}\text{, }r_{n}\right\} +\varepsilon _{ijk}\varepsilon _{lmn}r_{j}%
\mathring{\Gamma}_{n}\left\{ r_{m}\text{, }A_{k}\right\} \text{.}  \notag
\end{eqnarray}%
Finally, the fourth bracket on the rhs of (\ref{poi3a}) is given by%
\begin{equation}
\left\{ \varepsilon _{ijk}r_{j}A_{k}\text{, }\varepsilon
_{lmn}r_{m}A_{n}\right\} =-\varepsilon _{ijk}\varepsilon
_{lmn}r_{j}A_{n}\left\{ r_{m}\text{, }A_{k}\right\} -\varepsilon
_{ijk}\varepsilon _{lmn}r_{m}A_{k}\left\{ A_{n}\text{, }r_{j}\right\} \text{.%
}
\end{equation}%
Combining these results, we obtain%
\begin{eqnarray}
\left\{ J_{i}\text{, }J_{l}\right\}  &=&r_{l}p_{i}^{\text{flat}}-r_{i}p_{l}^{%
\text{flat}}+\delta _{il}r_{n}\mathring{\Gamma}_{n}-\delta _{il}r_{k}%
\mathring{\Gamma}_{k}+r_{i}\mathring{\Gamma}_{l}-r_{l}\mathring{\Gamma}_{i}+
\notag \\
&&+\delta _{il}r_{n}A_{n}-\delta _{il}r_{n}A_{n}+r_{l}A_{i}-r_{l}A_{i}+ 
\notag \\
&&+\varepsilon _{ijk}\varepsilon _{lmn}\left( r_{m}p_{k}^{\text{flat}%
}\left\{ A_{n}\text{, }r_{j}\right\} -r_{j}p_{n}^{\text{flat}}\left\{ r_{m}%
\text{, }A_{k}\right\} \right) +  \notag \\
&&+\varepsilon _{ijk}\varepsilon _{lmn}\left( r_{j}A_{n}\left\{ A_{k}\text{, 
}r_{m}\right\} -r_{m}A_{k}\left\{ A_{n}\text{, }r_{j}\right\} \right) + 
\notag \\
&&+\varepsilon _{ijk}\varepsilon _{lmn}\left( r_{m}p_{k}^{\text{flat}%
}\left\{ \mathring{\Gamma}_{n}\text{, }r_{j}\right\} -r_{j}p_{n}^{\text{flat}%
}\left\{ r_{m}\text{, }\mathring{\Gamma}_{k}\right\} \right) +  \notag \\
&&+\varepsilon _{ijk}\varepsilon _{lmn}\left( r_{j}\mathring{\Gamma}%
_{n}\left\{ \mathring{\Gamma}_{k}\text{, }r_{m}\right\} -r_{m}\mathring{%
\Gamma}_{k}\left\{ \mathring{\Gamma}_{n}\text{, }r_{j}\right\} \right) + 
\notag \\
&&+\varepsilon _{ijk}\varepsilon _{lmn}\left( r_{j}\mathring{\Gamma}%
_{n}\left\{ A_{k}\text{, }r_{m}\right\} -r_{m}\mathring{\Gamma}_{k}\left\{
A_{n}\text{, }r_{j}\right\} \right) +  \notag \\
&&+\varepsilon _{ijk}\varepsilon _{lmn}\left( r_{j}A_{n}\left\{ \mathring{%
\Gamma}_{k}\text{, }r_{m}\right\} -r_{m}A_{k}\left\{ \mathring{\Gamma}_{n}%
\text{, }r_{j}\right\} \right) -\varepsilon _{ilk}s_{k}\text{.}
\end{eqnarray}%
The full antisymmetry of the Levi-Civita tensor leads to%
\begin{eqnarray}
\varepsilon _{ijk}\varepsilon _{lmn}r_{m}p_{k}^{\text{flat}}\left\{ A_{n}%
\text{, }r_{j}\right\} -\varepsilon _{ijk}\varepsilon _{lmn}r_{j}p_{n}^{%
\text{flat}}\left\{ A_{n}\text{, }r_{m}\right\}  &=&\left( \varepsilon
_{ijk}\varepsilon _{lmn}-\varepsilon _{imn}\varepsilon _{ljk}\right)
r_{m}p_{k}^{\text{flat}}\left\{ A_{n}\text{, }r_{j}\right\}   \notag \\
&=&0
\end{eqnarray}%
and%
\begin{eqnarray}
\varepsilon _{ijk}\varepsilon _{lmn}r_{m}p_{k}^{\text{flat}}\left\{ 
\mathring{\Gamma}_{n}\text{, }r_{j}\right\} -\varepsilon _{ijk}\varepsilon
_{lmn}r_{j}p_{n}^{\text{flat}}\left\{ \mathring{\Gamma}_{n}\text{, }%
r_{m}\right\}  &=&\left( \varepsilon _{ijk}\varepsilon _{lmn}-\varepsilon
_{imn}\varepsilon _{ljk}\right) r_{m}p_{k}^{\text{flat}}\left\{ \mathring{%
\Gamma}_{n}\text{, }r_{j}\right\}   \notag \\
&=&0\text{.}
\end{eqnarray}%
Thus,%
\begin{equation}
\left\{ J_{i}\text{, }J_{l}\right\} =-\varepsilon _{ijk}J_{k}  \label{genrot}
\end{equation}%
\textbf{\ }proving that $\vec{J}$\ is the generator of spatial rotations.
Using $\vec{J}$\ we can now show that $e^{-2\phi \left( \left\vert \vec{r}%
\right\vert \right) }\mathcal{\vec{E}}\left( \vec{r}\right) $\textbf{\ }%
transforms as a vector,%
\begin{eqnarray}
\left\{ J_{i}\text{, }e^{-2\phi \left( \left\vert \vec{r}\right\vert \right)
}\mathcal{E}_{l}\right\}  &=&\left\{ J_{i}\text{, }e^{-2\phi \left(
\left\vert \vec{r}\right\vert \right) }\right\} \mathcal{E}_{l}+e^{-2\phi
\left( \left\vert \vec{r}\right\vert \right) }\left\{ J_{i}\text{, }\mathcal{%
E}_{l}\right\}   \label{evect} \\
&=&\varepsilon _{ilk}e^{-2\phi \left( \left\vert \vec{r}\right\vert \right) }%
\mathcal{E}_{k}\text{ since }\left\{ J_{i}\text{, }e^{-2\phi \left(
\left\vert \vec{r}\right\vert \right) }\right\} =0\text{.}  \notag
\end{eqnarray}%
\textbf{\ }The vector $e^{-2\phi \left( \left\vert \vec{r}\right\vert
\right) }\mathcal{\vec{E}}\left( \vec{r}\right) $ can be shown \cite{joshi}
to have a general functional form%
\begin{equation}
e^{-2\phi \left( \left\vert \vec{r}\right\vert \right) }\mathcal{\vec{E}}%
\left( \vec{r}\right) =k_{0}m_{\mathcal{E}}^{2}\xi \left( m_{\mathcal{E}}r%
\text{, }m_{\mathcal{E}}\vec{r}\cdot \hat{n}\right) (\hat{n}\times \vec{r})%
\text{,}  \label{abprime}
\end{equation}%
where $\xi $\ is a generic scalar field function. We emphasize again that
form the $\vec{\partial}\times \vec{B}^{\prime }\left( \vec{r}\right) $\
equation in (\ref{temp}), it is evident that no spherically symmetric
magnetic $\vec{B}^{\prime }\left( \vec{r}\right) $\ field solution exists,
i.e. $\vec{B}^{\prime }\left( \vec{r}\right) \neq B^{\prime }\left( r\right) 
\hat{r}$.

We now study whether the symmetry properties of the magnetic field obtained
above is compatible with the Poisson algebra of the system. This is
accomplished by determining if the magnetic field transforms as a vector
under spatial rotations by computing the Poisson bracket $\left\{ J_{i}\text{%
, }B_{j}\right\} $. It is convenient to begin this analysis by calculating
the curved space canonical momentum $\vec{p}$\ and the curved space kinetic
momentum vector $\vec{P}$\ Poisson brackets,\textbf{\ }%
\begin{eqnarray}
\left\{ p_{i}\text{, }p_{j}\right\} &=&\left\{ p_{i}^{\text{flat}}-\mathring{%
\Gamma}_{i}\text{, }p_{j}^{\text{flat}}-\mathring{\Gamma}_{j}\right\}  \notag
\\
&=&\left\{ p_{i}^{\text{flat}}\text{, }p_{j}^{\text{flat}}\right\} -\left\{
p_{i}^{\text{flat}}\text{, }\mathring{\Gamma}_{j}\right\} -\left\{ \mathring{%
\Gamma}_{i}\text{, }p_{j}^{\text{flat}}\right\} +\left\{ \mathring{\Gamma}%
_{i}\text{, }\mathring{\Gamma}_{j}\right\}  \notag \\
&=&\left\{ \mathring{\Gamma}_{j}\text{, }p_{i}^{\text{flat}}\right\}
-\left\{ \mathring{\Gamma}_{i}\text{, }p_{j}^{\text{flat}}\right\} =-%
\mathfrak{R}_{ij}\text{where\textbf{\ }}\mathfrak{R}_{ij}=\partial _{i}%
\mathring{\Gamma}_{j}+\partial _{j}\mathring{\Gamma}_{i}  \label{poi1a}
\end{eqnarray}%
and%
\begin{eqnarray}
\left\{ P_{i}\text{, }P_{j}\right\} &=&\left\{ p_{i}-A_{i}\text{, }%
p_{j}-A_{j}\right\}  \label{poi2a} \\
&=&\left\{ p_{i}\text{, }p_{j}\right\} -\left\{ p_{i}\text{, }A_{j}\right\}
-\left\{ A_{i}\text{, }p_{j}\right\} +\left\{ A_{i}\text{, }A_{j}\right\} 
\notag \\
&=&\left\{ A_{j}\text{, }p_{i}\right\} -\left\{ A_{i}\text{, }p_{j}\right\} -%
\mathfrak{R}_{ij}  \notag \\
&=&\left\{ A_{j}\text{, }p_{i}^{\text{flat}}-\mathring{\Gamma}_{i}\right\}
-\left\{ A_{i}\text{, }p_{j}^{\text{flat}}-\mathring{\Gamma}_{j}\right\} -%
\mathfrak{R}_{ij}  \notag \\
&=&\left\{ A_{j}\text{, }p_{i}^{\text{flat}}\right\} -\left\{ A_{j}\text{, }%
\mathring{\Gamma}_{i}\right\} -\left\{ A_{i}\text{, }p_{j}^{\text{flat}%
}\right\} +\left\{ A_{i}\text{, }\mathring{\Gamma}_{j}\right\} -\mathfrak{R}%
_{ij}  \notag \\
&=&-\partial _{i}A_{j}+\partial _{j}A_{i}-\left\{ A_{j}\text{, }\mathring{%
\Gamma}_{i}\right\} +\left\{ A_{i}\text{, }\mathring{\Gamma}_{j}\right\} -%
\mathfrak{R}_{ij}  \notag \\
&=&-(\partial _{i}A_{j}-\partial _{j}A_{i})-\mathfrak{R}_{ij}=-\varepsilon
_{ijk}B_{k}-\mathfrak{R}_{ij}\text{.}  \notag
\end{eqnarray}%
Note that we employed the Dirac-veto $B_{k}=\left( \varepsilon
_{klm}\partial _{l}A_{m}+e_{\left( m\right) }h_{k}\right) \overset{\text{veto%
}}{\rightarrow }\varepsilon _{klm}\partial _{l}A_{m}$\ in obtaining (\ref%
{poi2a}), that is, we impose that the electrically charged particle must
never pass through the string \cite{Brandtand}\ and therefore the electric
charge does not "feel" the magnetic field contribution originating from the
string function $\vec{h}\left( \vec{r}\right) $. From (\ref{poi2a}) we
conclude%
\begin{equation}
B_{k}=-\frac{1}{2}\varepsilon _{ijk}\left( \left\{ P_{i}\text{, }%
P_{j}\right\} +\mathfrak{R}_{ij}\right) \text{.}
\end{equation}%
We can now calculate the Poisson bracket $\left\{ J_{i}\text{, }%
B_{j}\right\} $ 
\begin{eqnarray}
\left\{ J_{i}\text{, }B_{j}\right\} &=&-\frac{1}{2}\varepsilon _{lmj}\left\{
J_{i}\text{, }\mathfrak{R}_{lm}+\left\{ P_{l}\text{, }P_{m}\right\} \right\}
\label{jbPoisson} \\
&=&-\frac{1}{2}\varepsilon _{lmj}\left\{ J_{i}\text{, }\mathfrak{R}%
_{lm}\right\} -\frac{1}{2}\varepsilon _{lmj}\left\{ J_{i}\text{, }\left\{
P_{l}\text{, }P_{m}\right\} \right\}  \notag \\
&=&\frac{1}{2}\varepsilon _{lmj}\left\{ J_{i}\text{, }\left\{ p_{l}\text{, }%
p_{m}\right\} \right\} -\frac{1}{2}\varepsilon _{lmj}\left\{ J_{i}\text{, }%
\left\{ P_{l}\text{, }P_{m}\right\} \right\}  \notag
\end{eqnarray}%
by using the Jacobi identities%
\begin{equation}
\left\{ J_{i}\text{, }\left\{ P_{l}\text{, }P_{m}\right\} \right\} +\left\{
P_{m}\text{, }\left\{ J_{i}\text{, }P_{l}\right\} \right\} +\left\{ P_{l}%
\text{, }\left\{ P_{m}\text{, }J_{i}\right\} \right\} =0\text{,}
\label{jcby1}
\end{equation}%
\begin{equation}
\left\{ J_{i}\text{, }\left\{ p_{_{l}}\text{, }p_{m}\right\} \right\}
+\left\{ p_{m}\text{, }\left\{ J_{i}\text{, }p_{_{l}}\right\} \right\}
+\left\{ p_{_{l}}\text{, }\left\{ p_{m}\text{, }J_{i}\right\} \right\} =0%
\text{,}  \label{jcby2}
\end{equation}%
and 
\begin{equation}
\left\{ P_{m}\text{, }\left\{ J_{i}\text{, }P_{l}\right\} \right\}
=-\varepsilon _{ilk}\left\{ P_{m}\text{, }P_{k}\right\} =-\varepsilon _{ilk}%
\left[ -\varepsilon _{mkn}B_{n}-\mathfrak{R}_{mk}\right] \text{,}
\label{temp1}
\end{equation}%
\begin{equation}
\left\{ P_{l}\text{, }\left\{ P_{m}\text{, }J_{i}\right\} \right\} =-\left\{
P_{l}\text{, }\left\{ J_{i}\text{, }P_{m}\right\} \right\} =\varepsilon
_{imk}\left\{ P_{l}\text{, }P_{k}\right\} =\varepsilon _{imk}\left[
-\varepsilon _{lkn}B_{n}-\mathfrak{R}_{lk}\right] \text{,}  \label{temp2}
\end{equation}%
\begin{equation}
\left\{ p_{m}\text{, }\left\{ J_{i}\text{, }p_{l}\right\} \right\}
=-\varepsilon _{ilk}\left\{ p_{m}\text{, }p_{k}\right\} =-\varepsilon
_{ilk}\left( -\mathfrak{R}_{mk}\right) \text{,}  \label{temp3}
\end{equation}%
\begin{equation}
\left\{ p_{l}\text{, }\left\{ p_{m}\text{, }J_{i}\right\} \right\} =-\left\{
p_{l}\text{, }\left\{ J_{i}\text{, }p_{m}\right\} \right\} =\varepsilon
_{imk}\left\{ p_{l}\text{, }p_{k}\right\} =\varepsilon _{imk}\left( -%
\mathfrak{R}_{lk}\right) \text{.}  \label{temp4}
\end{equation}%
Using (\ref{temp1}), (\ref{temp2}), (\ref{temp3}), (\ref{temp4}) together
with the Jacobi identities (\ref{jcby1}) and (\ref{jcby2}), we obtain 
\begin{eqnarray}
\left\{ J_{i}\text{, }\left\{ P_{l}\text{, }P_{m}\right\} \right\}
&=&-\left( -\varepsilon _{ilk}\left[ -\varepsilon _{mkn}B_{n}-\mathfrak{R}%
_{mk}\right] +\varepsilon _{imk}\left[ -\varepsilon _{lkn}B_{n}-\mathfrak{R}%
_{lk}\right] \right)  \label{bra1} \\
&=&-\varepsilon _{ilk}\varepsilon _{mkn}B_{n}+\varepsilon _{imk}\varepsilon
_{lkn}B_{n}-\varepsilon _{ilk}\mathfrak{R}_{mk}+\varepsilon _{imk}\mathfrak{R%
}_{lk}  \notag \\
&=&-\delta _{il}B_{m}+\delta _{im}B_{l}-\varepsilon _{ilk}\mathfrak{R}%
_{mk}+\varepsilon _{imk}\mathfrak{R}_{lk}  \notag
\end{eqnarray}%
and%
\begin{eqnarray}
\left\{ J_{i}\text{, }\left\{ p_{l}\text{, }p_{m}\right\} \right\} &=&-\left[
-\varepsilon _{ilk}\left( -\mathfrak{R}_{mk}\right) +\varepsilon
_{imk}\left( -\mathfrak{R}_{lk}\right) \right]  \label{bra2} \\
&=&-\varepsilon _{ilk}\mathfrak{R}_{mk}+\varepsilon _{imk}\mathfrak{R}_{lk}%
\text{.}  \notag
\end{eqnarray}%
Substituting (\ref{bra1}) and (\ref{bra2}) into (\ref{jbPoisson}) leads to%
\begin{eqnarray}
\left\{ J_{i}\text{, }B_{j}\right\} &=&\frac{1}{2}\varepsilon _{lmj}\left[
-\varepsilon _{ilk}\mathfrak{R}_{mk}+\varepsilon _{imk}\mathfrak{R}_{lk}%
\right] -\frac{1}{2}\varepsilon _{lmj}\left[ -\delta _{il}B_{m}+\delta
_{im}B_{l}-\varepsilon _{ilk}\mathfrak{R}_{mk}+\varepsilon _{imk}\mathfrak{R}%
_{lk}\right]  \label{jb} \\
&=&\frac{1}{2}\varepsilon _{lmj}\left( \delta _{il}B_{m}-\delta
_{im}B_{l}\right) +\frac{1}{2}\varepsilon _{lmj}\varepsilon _{imk}\left( 
\mathfrak{R}_{lk}-\mathfrak{R}_{lk}\right) +\frac{1}{2}\varepsilon
_{lmj}\varepsilon _{ilk}\left( \mathfrak{R}_{mk}-\mathfrak{R}_{mk}\right) 
\notag \\
&=&-\varepsilon _{mij}B_{m}\text{.}  \notag
\end{eqnarray}%
It is known from (\ref{mag-mon}) that $\vec{B}_{\text{monopole}}$ is
spherically symmetric and following \cite{cafali}, it can be shown that the
diffuse magnetic field (with vector potential of form (\ref{abprime})) must
exhibit spherical symmetry%
\begin{equation}
\vec{B}^{\prime }\left( \vec{r}\right) =B^{\prime }\left( r\right) \hat{r}%
\text{.}
\end{equation}%
in order to satisfy (\ref{jb}). Such spherically symmetric solutions
however, are incompatible with the second equation in (\ref{temp}). This
result implies it is not possible to formulate a consistent classical theory
describing nonrelativistic point-like charged particles interacting with
magnetic monopoles without a "visible" string via topologically massive
vector bosons in curved spacetime with isotropic dilation since there is no
way to construct a consistent Lie algebra.

\section{Conclusion}

In this article we considered a Brans-Dicke generalization of gravity with
non-vanishing curvature and torsion of potential type. An action describing
electromagnetic interaction between charged, nonrelativistic\textbf{\ }%
fermions with an abelian magnetic monopole, where the interaction is
mediated by topologically massive vector bosons, was proposed. The gauge
field mass is a direct consequence of the (topological) coupling -
characterized by $\mu _{0}$ - between the electromagnetic $4$-vector and the
second-rank torsion potential. This coupling is said to be topological due
to the lack of $\mu _{0}$-dependent terms in the canonical energy-momentum
tensor. The field equations for the theory as well as the Bianchi identities
in the electromagnetic and torsion sectors were obtained. From the solutions
to the torsion field equation (\ref{TorsionSoln}) we observe that the
sources of torsion are spinors, dilatons and photons. The dilatonic
contribution arises from the non-minimal torsion-dilaton coupling while the
electromagnetic contribution is due to the aforementioned topological
interaction.

Assuming an isotropic dilaton field configuration, the quantity\textbf{\ }$%
e^{-2\phi \left( \left\vert \vec{r}\right\vert \right) }\mathcal{\vec{E}}$%
\textbf{\ }plays the role of a massive photon-like term with mass $m_{%
\mathcal{E}}^{2}=2\mu _{0}$. This term together with pseudo-current $\vec{j}%
_{5}$\ arising from the spin energy potential constitute the total diffusive
magnetic potential $\vec{A}^{\prime }$. It was demonstrated that the Poisson
bracket $\left\{ J\text{, }B\right\} $\ in curved spacetime is not only well
defined but identical in structure to the flat spacetime dilaton free case.
It can be shown following \cite{cafali} that under the isotropic dilaton and
Dirac veto \textit{ansatz}, together with the limit\textbf{\ }$\left\vert 
\vec{j}_{5}\right\vert \ll \left\vert \frac{1}{k_{0}}e^{-2\phi \left(
\left\vert \vec{r}\right\vert \right) }\mathcal{\vec{E}}\left( \vec{r}%
\right) \right\vert $, spherically symmetric magnetic field solutions are
required in order to satisfy the Poisson bracket $\left\{ J\text{, }%
B\right\} $. Although $\vec{B}_{\text{monopole}}$\ is spherically symmetric,
spherical solutions for the diffuse magnetic field $\vec{B}^{\prime }$\ are
inconsistent with the nonvanishing of $\vec{\partial}\times \vec{B}^{\prime
}\left( \vec{r}\right) $\ in (\ref{temp}) despite the topological nature of
photon mass effectively generated by $m_{\mathcal{E}}^{2}e^{-2\phi \left(
\left\vert \vec{r}\right\vert \right) }\mathcal{\vec{E}}\left( \vec{r}%
\right) $. For this reason we conclude that the incompatibility between
massive photons and magnetic monopoles (without visible string) in the
present classical framework is not a consequence of the specific nature of
photon mass generation. What is more, the incompatibility survives the
transition from flat to curved spacetime and persists even in presence of
(isotropic) dilaton fields. With regard to the matter content of the theory,
it is interesting to observe that depending on the sign of the fermion
electric charge, the pseudo-current $\vec{j}_{5}$\ could serve to either
enhance or degrade the massive photon-like term. A measurable consequence of
this would be an associated increase or decrease of the diffuse magnetic
field intensity arising from the diffuse vector potential $\vec{A}^{\prime }$%
.

\section{Acknowledgements}

The authors thank an anonymous Referee for very useful comments that led to
significant, concrete improvements of this work.


\begin{thebibliography}{99}
\bibitem{GRG} S. Capozziello, M. Francaviglia, arXiv: gr-qc/0706.1146 (2007)

\bibitem{Dirac} P. A. M. Dirac, Proc. Roy. Soc. A133, 60 (1931); P. A. M.
Dirac, Phys. Rev. \textbf{74}, (1948) 817

\bibitem{Ogievetskii} V. I. Ogievetskii and I. V. Polubarinov, Sov. J. Nucl.
Phys. \textbf{4} (1967) 156

\bibitem{KalbRaymond} M. Kalb and P. Raymond, Phys. Rev. \textbf{D9} (1974)
2273

\bibitem{ScherkSchwarz} J. Scherk and J. H. Schwartz, Phys. Lett. \textbf{B52%
} (1974) 347

\bibitem{FradkinTseytlin} E. S. Fradkin and A. A. Tseytlin, Phys. Lett 
\textbf{B158} (1985) 316

\bibitem{BransDicke} C. H. Brans and R. H. Dicke, Phys. Rev. \textbf{124},
925 (1961)

\bibitem{Brans} C. H. Brans, \textit{1st International Workshop on
Gravitation and Cosmology}, Santa Clara, Cuba, 31 May - 4 June 2004, arXiv:
gr-qc/0506063 (2005)

\bibitem{Hehl1} F. W. Hehl \textit{et al}., Rev. Mod. Phys. \textbf{48}
(1976) 393

\bibitem{Kibble} T. W. Kibble, J. Math. Phys. \textbf{2} (1960) 212

\bibitem{cho2} Y. M. Cho, \textquotedblleft Gauge theory, gravitation and
symmetry\textquotedblright , Phys. Rev. \textbf{D14}, (1976) 3341; \textbf{"}%
Gauge theory of Poincar\'{e} symmetry", Phys. Rev. \textbf{D14}, (1976) 3335%
\textbf{.}

\bibitem{Blagojevic} M. Blagojevic, \textit{2nd Summer School in Modern
Mathematical Physics}, Kopaonik, Serbia, Yugoslavia, 1-12 Sep 2002.
Published in SFIN A1:147-172,2003, arXiv: gr-qc/0302040

\bibitem{Grignani} G. Grignani, Phys. Rev. \textbf{D45} (1992) 2719

\bibitem{Hammond1} R. T. Hammond, Class. Quantum Grav. \textbf{13} (1996)
L73; Class. Quantum Grav. \textbf{12} (1995) 279; Gen. Rel. Grav. \textbf{26}
247

\bibitem{noi} S. Capozziello, G. Lambiase, C. Stornaiolo, Ann. Phys.
(Leipzig) \textbf{10}, (2001) 713

\bibitem{joshi} A. Yu Ignatiev and G. C. Joshi, Phys. Rev. \textbf{D53}, 984
(1995)

\bibitem{cafali} C. Cafaro, S. Capozziello, Ch. Corda and S. A. Ali, Adv.
High Energy Phys. Article ID \textbf{69835} (2007)

\bibitem{Wald} R. M. Wald, \textit{Quantum field theory in curved spacetime
and black hole thermodynamics}, (The University of Chicago Press 1994)

\bibitem{BirrellDavies} N. D. Birrell and P. C. W. Davies, \textit{Quantum
fields in curved space}, (Cambrideg University Press 1982)

\bibitem{Carroll} S.M. Carroll, G.B. Field, Phys. Rev. \textbf{D50}, 3867
(1994)

\bibitem{Schouten} J. Schouten, \textit{Ricci Calculus} (Berlin: Springer,
1954)

\bibitem{crawford} J. P. Crawford, Spinors in General Relativity, appearing
in "Clifford (Geometric) Algebras: With Applications in Physics, Mathematics
and Engineering", Birkh\"{a}user, Boston (1996) edited by William E. Baylis.

\bibitem{cho} Y. M. Cho, \textquotedblleft Reinterpretation of
Jordan-Brans-Dicke Theory and Kaluza-Klein Cosmology\textquotedblright ,
Phys. Rev. Lett \textbf{68}, 3133 (1992)

\bibitem{Gasparini} M. Gasperini and R. Ricci, Class. Quantum Grav. \textbf{%
12} (1995) 677

\bibitem{Shapiro} I. L. Shapiro, Phys. Rept. \textbf{357}, 113, (2002)

\bibitem{Moura-Melo} W. A. Moura-Melo, N. Panza and J. A. Helayel-Neto, Int.
J. Mod. Phys. \textbf{A14}, 3949 (1999).

\bibitem{Brill} D. R. Brill and J. A. Wheeler, Rev. Mod. Phys. \textbf{29}
(1957) 465

\bibitem{Hehl2} F. W. Hehl and B. K. Datta, J. Math. Phys. \textbf{12}
(1971) 1334

\bibitem{goldhaber} A. S. Goldhaber, Role of Spin in the Monopole Problem,
Phys. Rev. \textbf{D40}, B1407 (1965)

\bibitem{Berard} A. Berard, Y. Grandati and H. Mohrbach, "Dirac monopole
with Feynman brackets", Phys. Lett. \textbf{A254} (1999) 133

\bibitem{Cap} Giuseppe Basini and Salvatore Capozziello, Int. J. Mod. Phys. 
\textbf{D15} (2006) 583

\bibitem{Brandtand} R. A. Brandtand and J. R. Primack, Avoiding "Dirac's
veto" in monopole theory, Phys. Rev. \textbf{D15}, 1798 (1976)
\end{thebibliography}
\end{document}